\documentclass[twocolumn,showpacs,preprintnumbers,amsmath,amssymb]{revtex4}
\usepackage{balance}

\usepackage{comment}
\usepackage{multirow}
\usepackage{float} 

\usepackage[unicode]{hyperref}
\usepackage[caption=false]{subfig}
\usepackage{graphicx}
\usepackage{epsfig}
\usepackage{xcolor}
\graphicspath{ {./images/} }
\usepackage{dcolumn}
\usepackage{bm}
\newcommand{\ms}[1]{\textcolor{red}{#1}}

\begin{document}

\title{Topology and Choice-Driven Percolation in Mediation-Driven Attachment Networks}
\title{Role of Choice and Entropy in Percolation on Mediation-Driven Attachment Networks}
\title{Exploring the Power of Choice and Topology in Percolation on MDA Networks}
\title{Percolation in Mediation-Driven Attachment Networks: Exploring the Power of Choice and Topology}
\title{\ms{Percolation in Mediation-Driven Attachment Networks: Exploring the Impact of Choice and Network Topology}}
\title{Interplay of choice and topology in percolation on mediation-driven attachment networks}


\author{ Nilomber Roy$^{1}$, M. M. B. Sheraj$^{1,2,3}$ and M. K. Hassan$^{1}$\\
\small
$^1$University of Dhaka, Department of Physics, Dhaka 1000, Bangladesh \\
$^2$BRAC University, Department of Mathematics and Natural Sciences, Dhaka 1212, Bangladesh\\
$^3$ University of Texas at Austin, Department of Physics, Austin, TX 78712, USA \\
}%
\date{\today}%

\begin{abstract}
We investigate bond percolation on mediation-driven attachment (MDA) networks under the generalized Achlioptas process, where $M>1$ candidate bonds are sampled and the one that minimizes the resulting cluster size is selected—the best-of-$M$ rule. This framework offers a systematic approach to investigate how network topology and choice mechanisms jointly shape percolation behavior. We analyze the effects of the degree exponent $\omega$ and the choice parameter $M$ on the critical point $t_c$ and the critical exponents $(\beta,\alpha,\gamma)$, which define universality classes and obey the Rushbrooke inequality $\alpha + 2\beta + \gamma \ge 2$. Using entropy, the order parameter, and their derivatives (representing specific heat and susceptibility respectively), we show that both $t_c$ and the universality class depend only weakly on $\omega$ but strongly on $M$, while the Rushbrooke inequality remains valid throughout. For $M=2$, the order parameter varies continuously without a clear order-disorder transition. By contrast, $M=3$ and $M=4$ display explosive percolation that still corresponds to a continuous phase transition, with $M=4$ producing a significantly sharper and clearer order-disorder transition. This sharpening is traced to an enhanced powder-keg effect at larger $M$, underscoring the entropic origin of explosive percolation.
\end{abstract}


\maketitle

\section{Introduction}

The concept of percolation was first introduced by Flory in 1941 in the context of cross-linking and polymer gelation \cite{ref.flory}. However, the first rigorous mathematical formulation of percolation was provided by Broadbent and Hammersley in 1957 to understand the motion of gas molecules as they navigate the maze of pores within the carbon granules used to fill a gas mask \cite{ref.broadbent}. In order to define percolation, we first need to select an underlying structure or skeleton, which could be a lattice or a network. Regardless of its form, this skeleton is composed of a set of sites or nodes connected by bonds or links in a specific configuration. Depending on whether the occupancy is applied to the sites or the bonds, percolation is classified as either site percolation or bond percolation. Despite its seemingly straightforward definition, the percolation problem remains analytically unsolved for most dimensions, specifically for $1 < d < \infty$, except in certain special cases \cite{ref.Stauffer}. Extensive numerical studies on regular lattices have provided a foundation for understanding percolation phenomena \cite{ref.xun_hao_ziff,ref.xun_hao_ziff2,ref.mitra_saha_sensharma}.

The first systematic study of network percolation was conducted by Erd\H{o}s and R\'enyi (ER) in 1959, where $N$ isolated nodes are sequentially linked at random \cite{ref.erdos}. In 2005, Ben-Naim and Krapivsky demonstrated that the ER random graph process could be mapped onto a kinetic framework, where the density of occupied bonds is parameterized as time \cite{ref.BenNaimKrapiv}. 
The progression of time is denoted by the relative link density $t=n/N$ where $n$ represents the total number of links added thus far in the process. As link density 
$t$ increases, small components progressively coalesce. At a critical point $t_c$,  
the largest cluster $s_{\rm max}$ shifts from microscopic scaling ($\sim \log N$) to macroscopic scaling ($\sim N$), signaling the emergence of a giant connected component.
Earlier, Fisher and Essam (1961), and later Kasteleyn and Fortuin (1969), showed that the relative size of the largest cluster, $P(t, N) = \frac{s_{\rm max}}{N}$, plays the role of an order parameter, analogous to magnetization in spin systems \cite{ref.fisher,ref.Kasteleyn}. This analogy firmly establishes percolation as a paradigm of thermodynamic phase transitions and critical phenomena. Since then, percolation theory has evolved into a rich and vibrant field, owing to its remarkable ability to capture the universal properties of diverse and seemingly unrelated systems \cite{ref.Stanley,ref.saberi}.

Most early studies of percolation focused on the random sequential addition of a single site or bond at each time step \cite{ ref.martins_plascak,ref.grassberger_1}. In 2009, Achlioptas {\it et al.} introduced a striking variant of this random process \cite{ref.Achlioptas, ref.raissa}. They asked: what if, instead of adding a single randomly selected link at each step, two candidate links are sampled and only the one that minimizes the resulting cluster size is added? This choice-driven mechanism, known as the Achlioptas process, was found to delay the onset of the percolation transition—a natural and expected consequence of suppressing the rapid growth of large clusters. Yet the most unexpected observation was that the order parameter $P$ exhibited an unusually abrupt transition at the critical point $t_c$,
so much sharper than in ordinary percolation that it was initially interpreted as a discontinuous (first-order) transition \cite{ref.ziff_1, ref.Costa_2, ref.souza, ref.cho_1, ref.ara}. By 2011, however, more detailed analyses had firmly established that the so-called explosive percolation (EP) transition is in fact continuous, albeit with pronounced finite-size effects resembling those of a first-order transition \cite{ref.da_Costa, ref.grassberger_2, ref.Lee, ref.Riordan, ref.Tian}.

Random percolation on ER networks differs fundamentally from percolation on spatially embedded lattices. When lattices exhibit scale-free properties, as in the case of the weighted planar stochastic lattice (WPSL) \cite{ref.hassan_njp}, random percolation behaves very differently from percolation on regular two-dimensional lattices. In fact, it belongs to a universality class that is distinct from that of regular two-dimensional lattices \cite{ref.hassan_mijan_1,ref.hassan_mijan_2}. Furthermore, when percolation occurs on a skeleton that is both scale-free and, at the same time, a network rather than a lattice, whether random or competitive, it displays remarkably rich and complex critical behavior \cite{ref.radicchi_2,ref.kadovic,ref.schwartz_barabasi,ref.jasch}. Cohen \textit{et al.} demonstrated that the critical exponents of random percolation are strongly influenced by the degree exponent $\omega$ of the underlying power-law degree distribution. Specifically, for $\omega \leq 3$, the network becomes hub-dominated, resulting in an essentially vanishing percolation threshold ($t_c \to 0$) and exhibiting extreme resilience. In the intermediate regime $3 < \omega < 4$, the system departs from mean-field predictions and displays non–mean-field critical behavior. For $\omega > 4$, however, the system recovers classical mean-field characteristics \cite{ref.cohen_havlin}.
Building on this, Radicchi and Fortunato showed that in explosive percolation the nature of the transition also depends on $\omega$: it remains continuous for $\omega \leq 3$ but becomes discontinuous once $\omega > 3$ \cite{ref.radicchi_1}.

Inspired by the construction process of the 
WPSL, Hassan \textit{et al.} (2017) proposed the mediation-driven attachment (MDA) network model \cite{ref.hassan_liana, ref.hassan_liana_debashish}. This model captures the intuitive idea of preferential attachment, similar to the Barabási–Albert (BA) model, but in disguised form akin to the WPSL. In the MDA network, at each time step, a new node with $m$ links joins the network. It first selects a mediator node uniformly at random and then connects to $m$ randomly chosen neighbors of that mediator. This mechanism produces a power-law degree distribution with an exponent $\omega$ that explicitly depends on $m$, strongly for small $m$ and weakly for large $m$ such that $\omega \to 3$ as $m \to \infty$. This is in sharp contrast to the BA model, where $\omega = 3$ is fixed for all $m$ \cite{ref.barabasi}. Remarkably, Hassan \textit{et al.} demonstrated that the MDA rule becomes \textit{super-preferential} when $m$ is very small, particularly at $m = 1$, resulting in a ``winner-takes-all’’ effect where a single node dominates. As $m$ increases, this extreme preferential effect diminishes, and in the large-$m$ limit, the network converges to a classical scale-free BA like network.

In this article, we investigate the {\it best-of-$M$} bond percolation model on the MDA network, focusing on representative values $m = 50, 100, 200$, where the ``winner-takes-all’’ effect is effectively mitigated, using the generalized Achlioptas process.
At each step, $M>1$ candidate bonds are sampled, and the bond whose occupation minimizes the resulting cluster size is selected for occupation, while the remaining bonds are returned to the pool for subsequent sampling. This setting provides a systematic framework for examining how the degree exponent $\omega$, determined by the network-architecture parameter $m$, and the choice parameter $M$ together shape the nature and characteristics of the percolation transition.
The percolation transition is typically characterized by the critical point $t_c$ and critical exponents $\alpha$, $\beta$, and $\gamma$ which govern the specific heat, order parameter, and susceptibility respectively ~\cite{ref.ziff_2,ref.Bastas,ref.cho_2}. These exponents cannot just assume arbitrarily any value as they are locked by some relations like Rushbrooke inequality $\alpha + 2\beta + \gamma \geq 2$ which becomes equality under static scaling \cite{ref.Stanley} and determine the universality classes \cite{ref.hassan_didar}. 
We explicitly show that both $t_c$ and the universality class vary with $m$ and $M$, and that the Rushbrooke inequality holds in all cases. Compared with ER and BA networks, MDA networks exhibit distinct universality \cite{ref.sabbir,ref.digonto}. Notably, for $M=2$ the transition is never explosive for any $m$, while larger $M$ sharpens the transition: both $M=3$ and $M=4$ yield continuous explosive percolation, with $M=4$ significantly implying that the order-disorder transition is sharper. We attribute the entropic origin of the sharper transition in the {\it best-of-$M$} model to the stronger powder-keg effect that emerges at higher values of $M$.

The remainder of this article is organized as follows. Section II introduces the construction of the MDA skeleton and its degree distribution. In Section III, we examine the power of choice through the {\it best-of-$M$} Achlioptas process, which gives rise to explosive percolation. The finite-size scaling hypothesis is employed in Section IV to extract the critical exponents. 
Section V analyzes how network topology and choice shape the universality class, and Section VI concludes with a summary of our findings.

\section{Network Topology}
The construction of the mediation-driven attachment (MDA) network starts with a seed network of say $m_0 = 6$ nodes forming a complete graph (clique) and we label the nodes as $i = 1, 2, \dots, 6$. From then on a new node with 
$m<m_0$ links is added at each time according to the following algorithm:
\begin{itemize}
 \item \textbf{Mediator Selection:} Randomly select one node from the existing network with uniform probability and regard it as the \emph{mediator}.

    \item \textbf{Neighbor Selection:}  From the set of neighbors of the mediator, randomly choose $m$ distinct nodes.
   
    \item \textbf{Attachment:} Connect the new node to these $m$ chosen neighbors.
    
    \item \textbf{Iteration:} Repeat the above steps until the network reaches the desired size.
  \end{itemize}

\begin{figure}[h]
\centering
\subfloat[]
{
\includegraphics[height=3.5 cm, width=4.2 cm, clip=true, angle=0]
{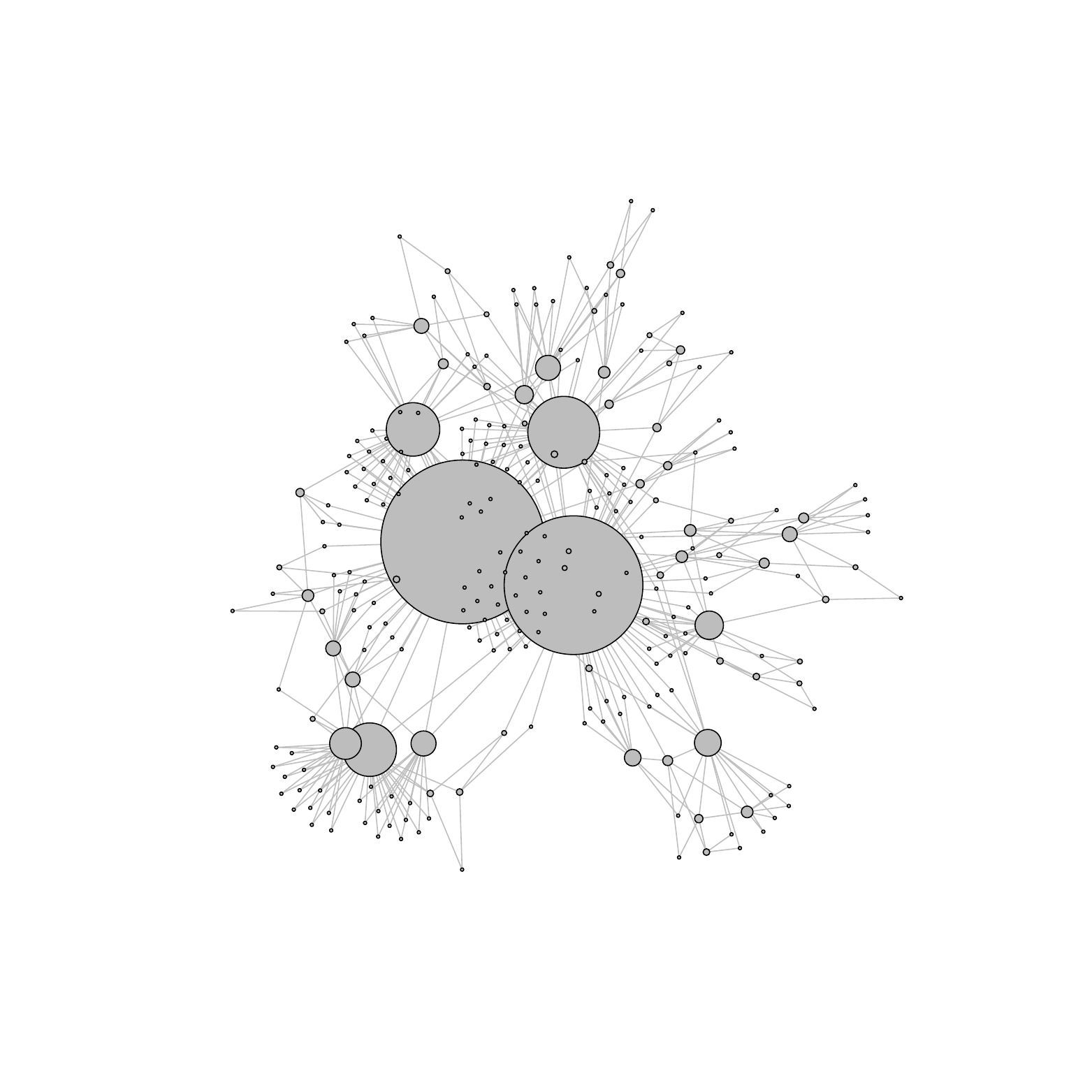}
\label{fig:1a}
}
\subfloat[]
{
\includegraphics[height=3 cm, width=4.2 cm, clip=true, angle=0]
{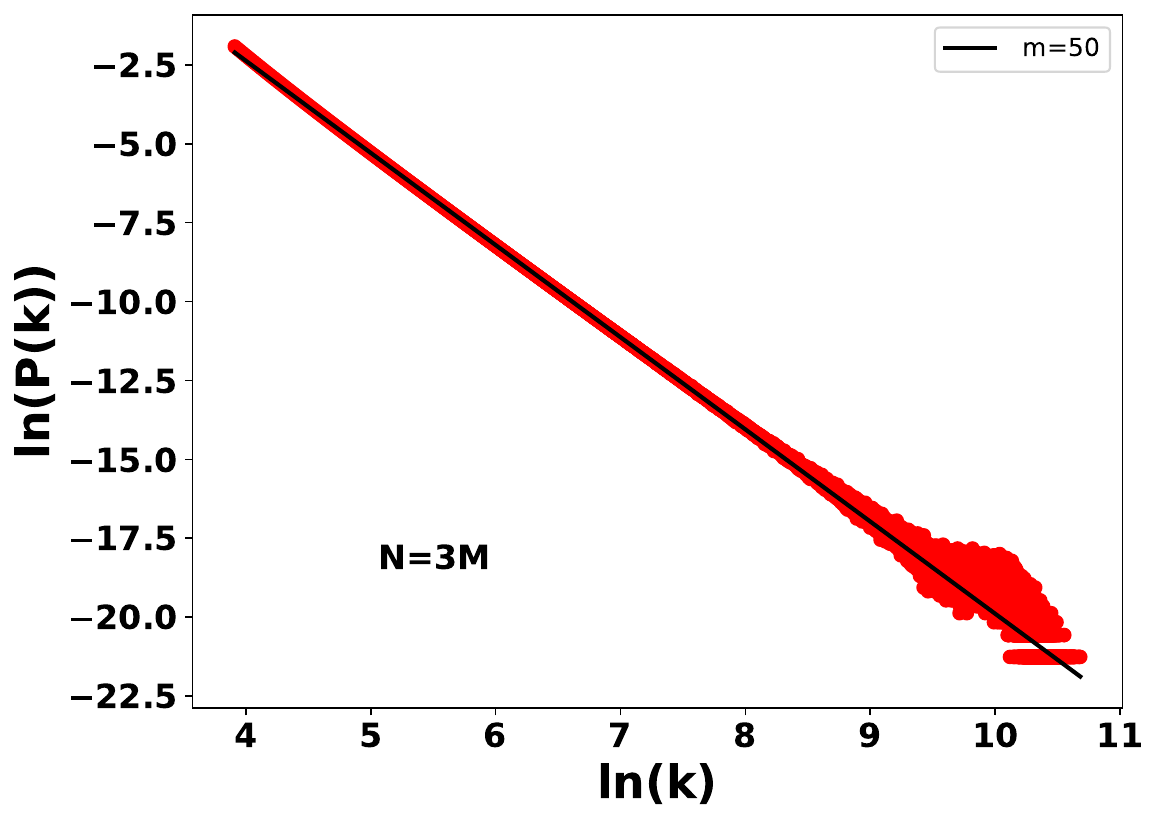}
\label{fig:1b}
}
\caption{We 
show a mineature of MDA network in (a). Next, plot (b) shows $\ln P(k)$ vs $\ln k$ for $m=50$, yielding a straight line with slope 
$2.9358$, representing the degree exponent.} 
\label{fig:1ab}
\end{figure}
The question is: how does the MDA network embody the intuitive idea of the preferential attachment rule? Note that the higher the number of nearest neighbors (i.e., the degree $k$) a node possesses, the greater the likelihood that one of its neighbors will be selected at random, thereby increasing its chances of forming a connection with the new node. Thus, the MDA network follows the preferential attachment principle, albeit indirectly or in disguise. This is evidenced by the emergence of a power-law degree distribution in the MDA network, consistent with Barab'{a}si's observation that growth and preferential attachment are the key mechanisms behind the scale-free nature of networks. However, unlike the BA network, the topology of the MDA network depends on the parameter $m$. For small values of $m$, particularly $m = 1$ and $m = 2$, a point in the $\ln P(k)$ vs. $\ln k$ plot appears significantly above the rest. This occurs because approximately $99.5\%$ of nodes have degree $k = 1$ for $m = 1$, and around $95.4\%$ have $k = 2$ for $m = 2$. These observations indicate that most nodes are weakly connected, while the network is predominantly held together by a few highly connected hubs, reflecting the so-called ``winner-takes-all'' effect~\cite{ref.hassan_liana}.
However, as $m$ value increases, the diminishing influence of these central nodes (i.e., hubs) leads to remarkable changes in the overall network architecture and connectivity patterns. 
Indeed, for $m=50$, the degree distribution follows a power-law decay $P(k)\sim k^{-\omega}$ with exponent $\omega=2.9358$, as shown in Fig.~\ref{fig:1ab}, while for $m=100$ and $m=200$, the corresponding exponents are $\omega=2.9808$ and $\omega=2.9989$, respectively.

\section{Power of choice in percolation}
Once the network of desired choice and size is grown, we use that as a skeleton for percolation.
To investigate the detailed nature of the percolation transition on these networks, we use the Achlioptas process, a competitive mechanism well known for producing explosive percolation, and analyze its behavior for $M=2,3,4$. 
\subsection{Achlioptas process}
Initially, we disconnect all links in the network and label them in a manner that records which pair of nodes each link connects. For instance, a link $e_{i,j}$ indicates a connection between two distinct nodes $i$ and $j$. Following the bond percolation rule, we select one link from all labeled links according to the {\it best-of-$M$} model and occupy it. We record the cluster sizes of the two nodes connected by the link, denoted by $s_i$ and $s_j$, respectively. If the occupation of a link $e_{i,j}$ connects two distinct clusters, the resulting cluster size becomes $s_i + s_j$, and the link is called an \emph{inter-cluster link}. Conversely, if a link connects two nodes already belonging to the same cluster, it does not increase the cluster size, and the link is termed an \emph{intra-cluster link}. Initially, each node forms a cluster of its own size, as all links are assumed unoccupied.

The process of competitive percolation begins by randomly selecting $M$ distinct candidate links from all unoccupied links with uniform probability at each step. The link to occupy is then chosen according to specific selection rules. Under the \emph{product rule}, the link that minimizes the product of the sizes of the clusters it connects is selected, whereas under the \emph{sum rule}, the link that minimizes the sum of the cluster sizes is chosen. To illustrate the product rule of the original Achlioptas process, we consider $M=2$ case. At each step, two unoccupied links, say $e_{ij}$ and $e_{kl}$, are selected, and the sizes of the clusters to which the nodes belong—$s_i$, $s_j$, $s_k$, and $s_l$—are recorded, 
where nodes $i$, $j$, $k$, and $l$ belong to these respective clusters. We then calculate the products
\begin{equation}
\Pi_{ij} = s_i \times s_j, \quad \text{and} \quad \Pi_{kl} = s_k \times s_l,
\end{equation}
noting that $s_i = s_j$ if nodes $i$ and $j$ belong to the same cluster. We then occupy the link $e_{ij}$ if $\Pi_{ij} < \Pi_{kl}$; otherwise, we occupy the link $e_{kl}$. In cases where $\Pi_{ij} = \Pi_{kl}$, one of the two links is selected at random with uniform probability. In all cases, the link that is not chosen is recycled for potential selection in future steps. In the case of $M=3$, we pick three links, say $e_{ij}$, $e_{kl}$ and
 $e_{mn}$, we then occupy the one that corresponds to the lowest value among $\Pi_{ij}$, $\Pi_{kl}$ and $\Pi_{mn}$. This idea is readily generalizable to any value of $M$. In this study, we use the product rule in edge selection process. For the simulations, we employ the Newman–Ziff (NZ) algorithm due to its superior computational efficiency in modeling percolation, and we apply their convolution technique to generate smooth curves for both the observables and their derivatives \cite{ref.newmanziff}. To ensure statistical robustness, we carry out $10^4$ independent realizations for each network size.

\subsection{Order--disorder transition}

\begin{figure}[h]
\centering

\subfloat[]
{
\includegraphics[height=3 cm, width=4.2 cm, clip=true, angle=0]
{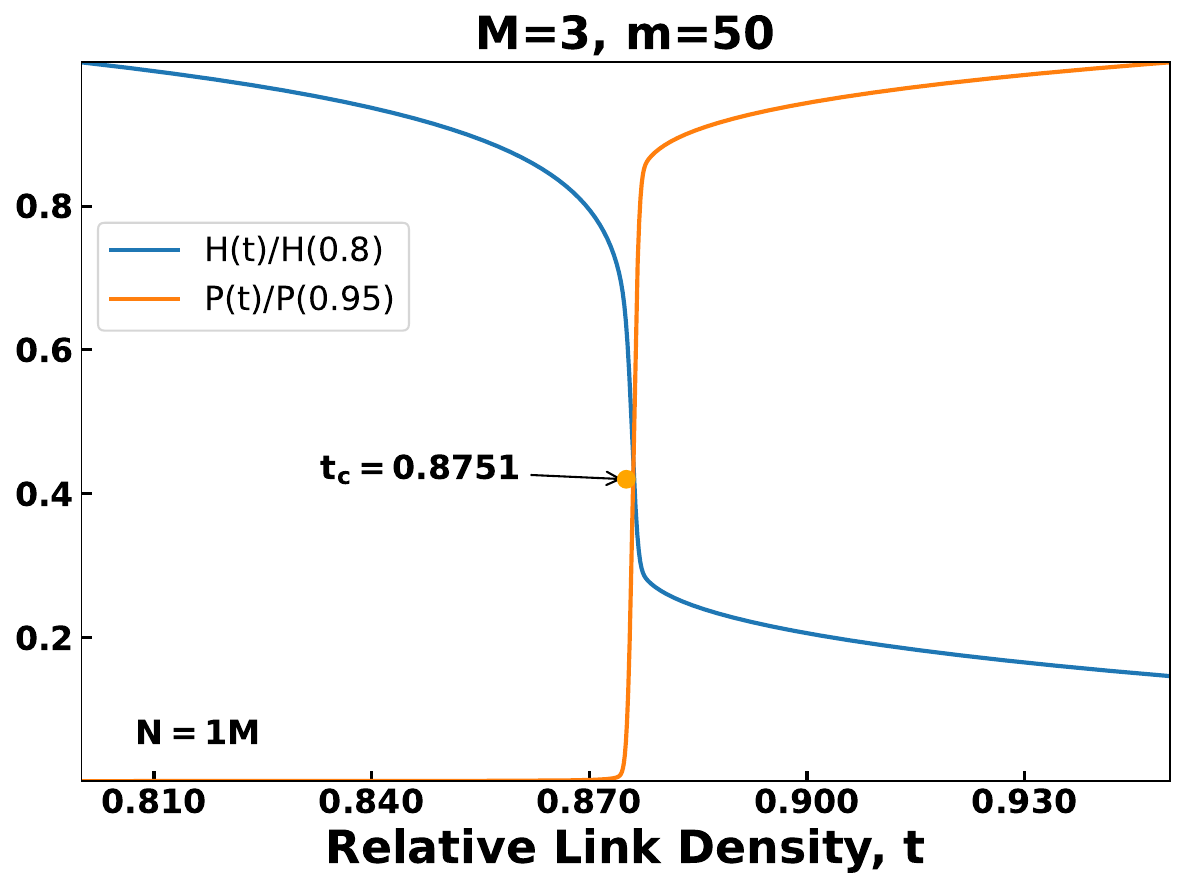}
\label{fig:2a}
}
\subfloat[]
{
\includegraphics[height=3 cm, width=4.2 cm, clip=true, angle=0]
{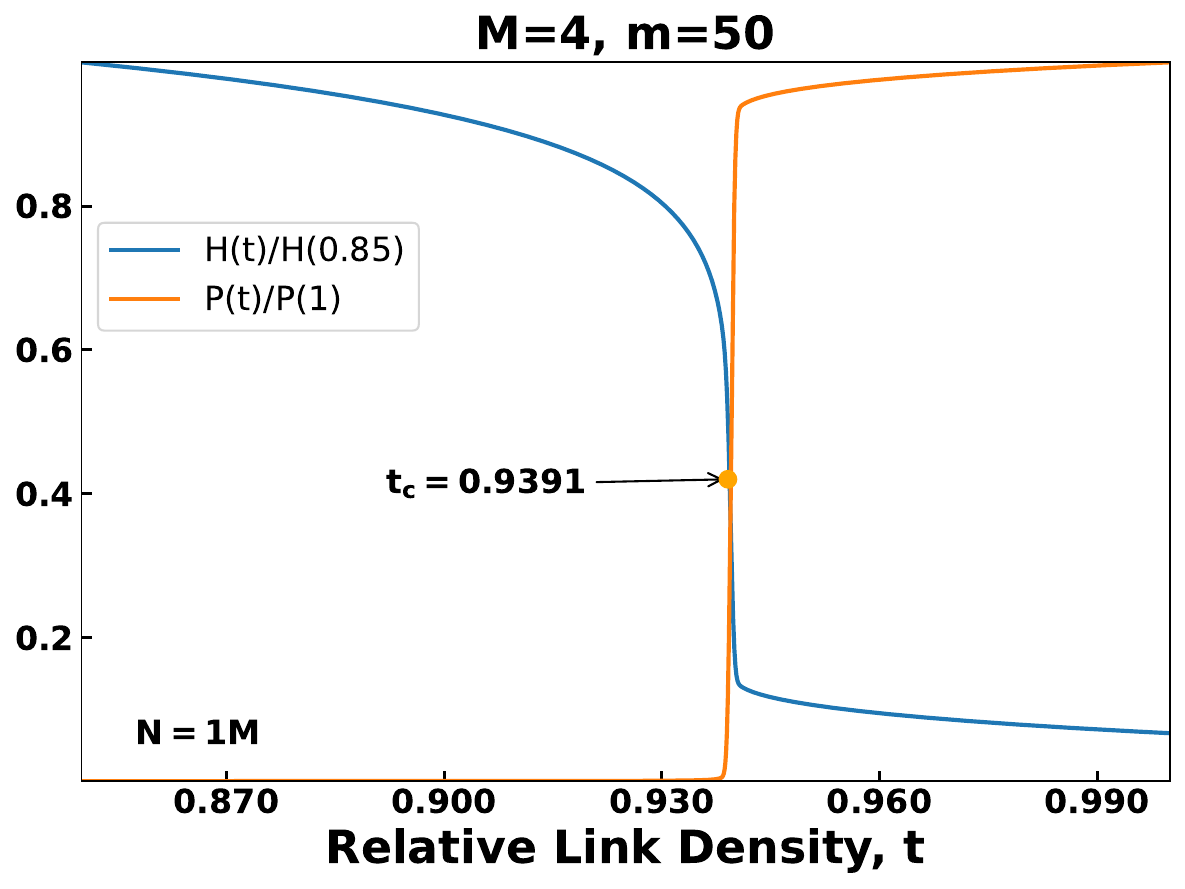}
\label{fig:2b}
}

\subfloat[]
{
\includegraphics[height=3 cm, width=4.2 cm, clip=true, angle=0]
{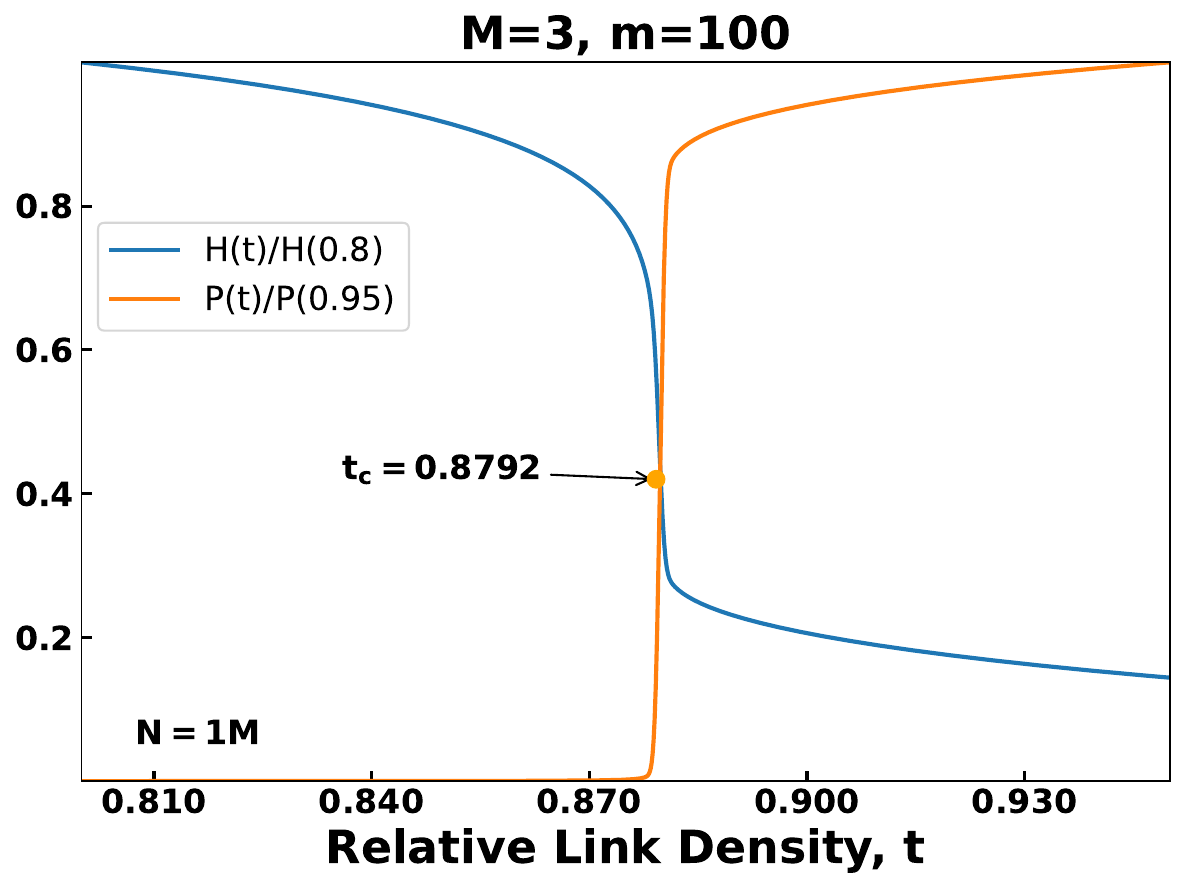}
\label{fig:2c}
}
\subfloat[]
{
\includegraphics[height=3 cm, width=4.2 cm, clip=true, angle=0]
{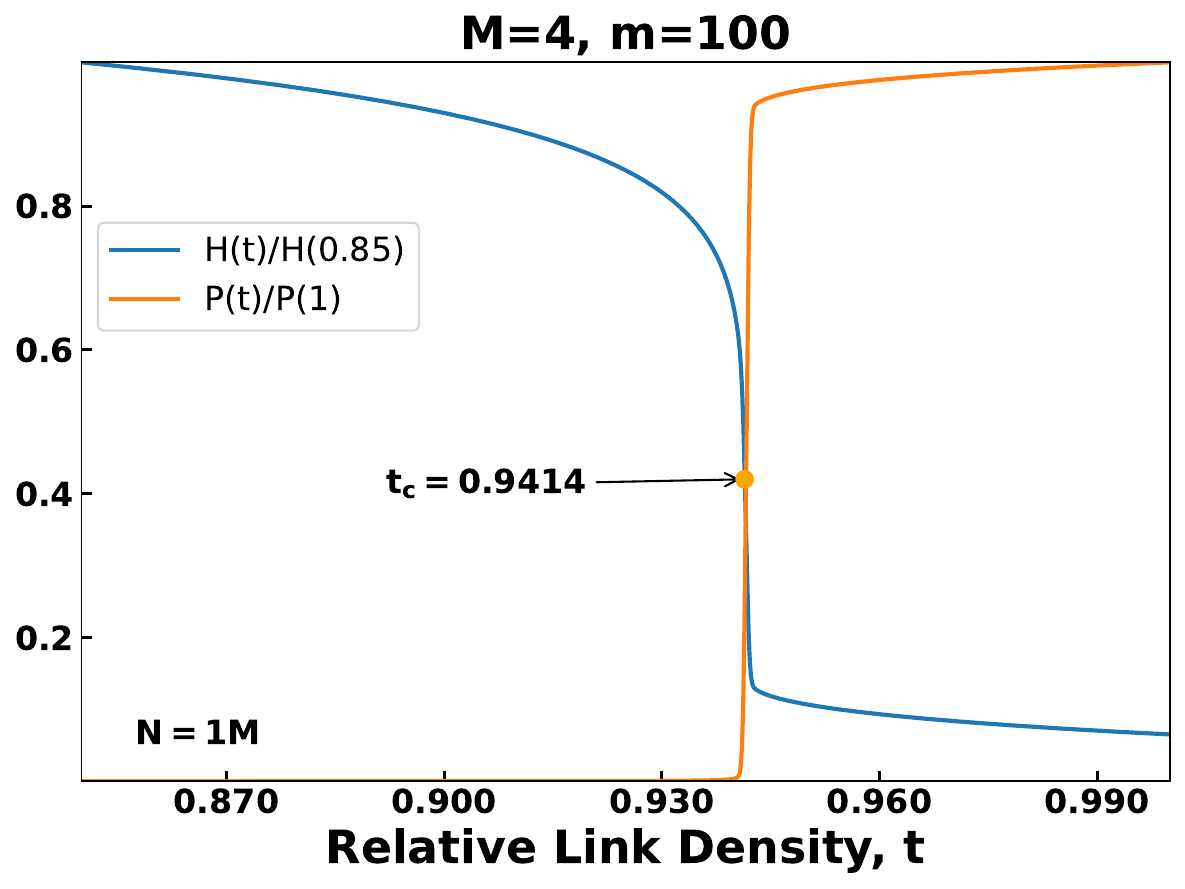}
\label{fig:2d}
}

\subfloat[]
{
\includegraphics[height=3 cm, width=4.2 cm, clip=true, angle=0]
{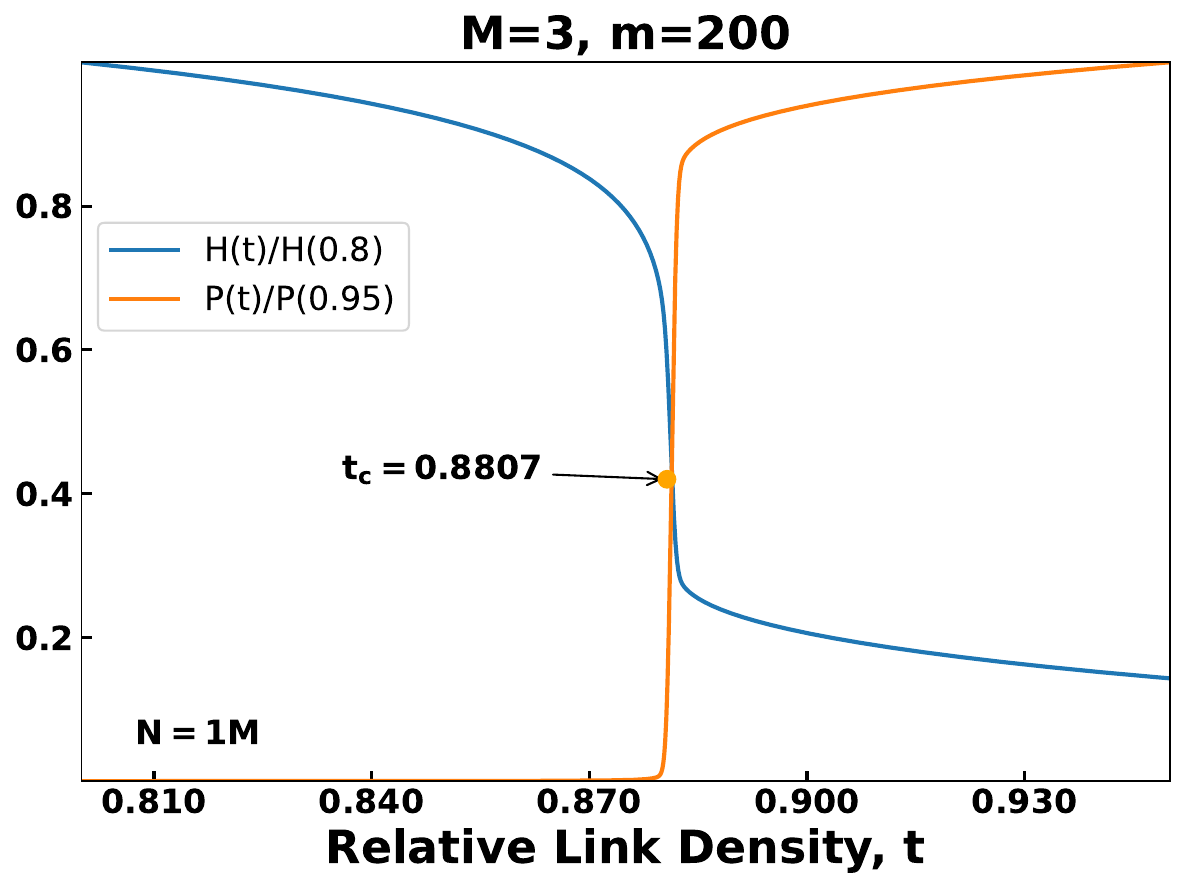}
\label{fig:2e}
}
\subfloat[]
{
\includegraphics[height=3 cm, width=4.2 cm, clip=true, angle=0]
{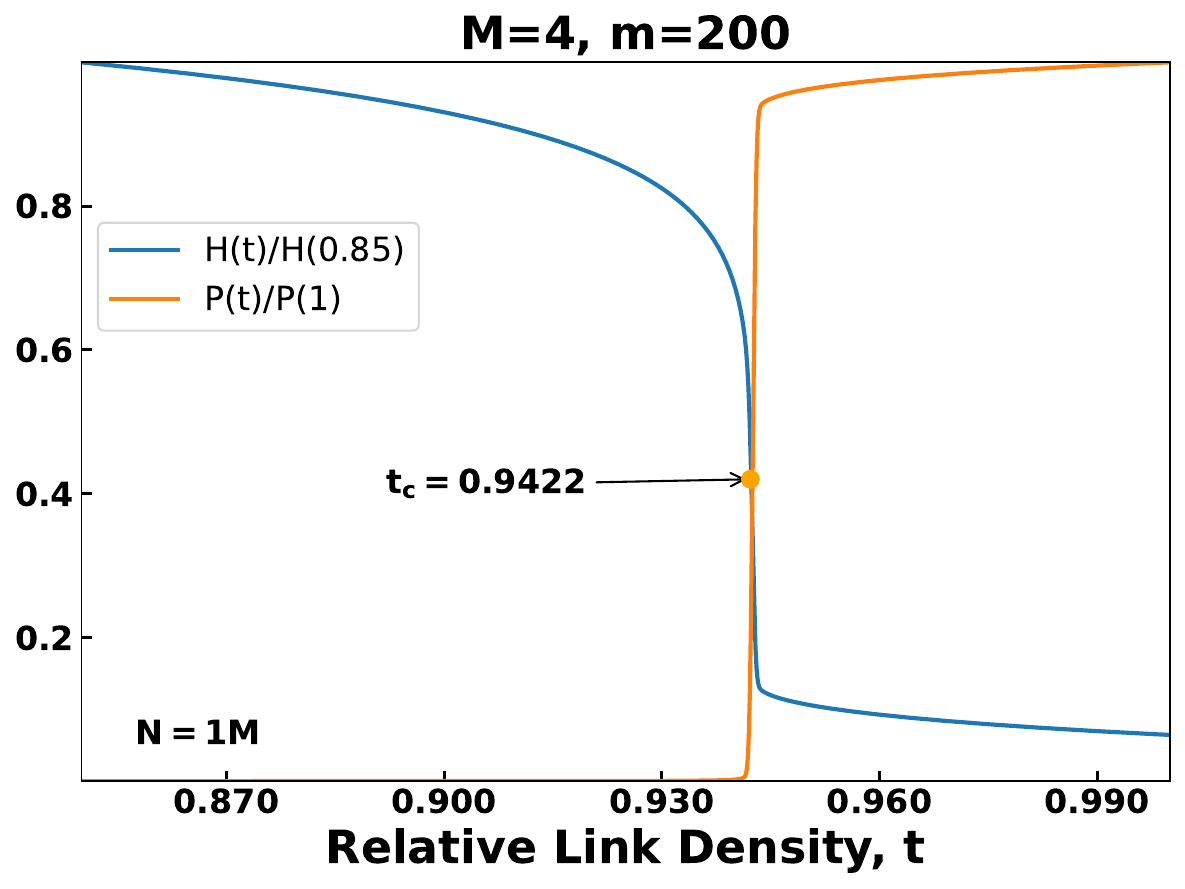}
\label{fig:2f}
}

\caption{Plots of relative entropy and relative order parameter for (a) $m=50,M=3$ (b) $m=50,M=4$ (c) $m=100,M=3$, (d) $m=100, M=4$ (e) $m=200,M=3$ and for (f) $m=200,M=4$.} 

\label{fig:2abcdef}
\end{figure}
In 1969, Fortuin and Kasteleyn (FK) introduced the random cluster representation of the $q$-state Potts model, showing that in the limit $q \rightarrow 1$, this representation reduces to bond percolation on a lattice~\cite{ref.Kasteleyn}. Within this framework, the relative size of the largest cluster,
\begin{equation}
    P = \frac{s_{\rm max}}{N},
\end{equation}
naturally emerges as the order parameter, analogous to the magnetization in thermal phase transitions, thereby allowing percolation to be studied within the general formalism of critical phenomena. While the order parameter $P$ effectively captures the degree of order in the ordered phase, it conveys no information about the disordered phase, since $P = 0$ throughout that entire regime. To gain a more comprehensive understanding of the percolation transition, it is therefore essential to also consider entropy, which quantifies the degree of disorder. It should be noted, however, that entropy need not remain constant across the whole region where $P=0$. Despite its importance, entropy in the context of percolation remained largely unexplored for more than six decades. The first notable attempts were made in the late 1990s by Tsang \textit{et al.}, followed by Vieira \textit{et al.} in 2015~\cite{ref.tsang, ref.vieira}.
\begin{figure}[h]
\centering

\subfloat[]
{
\includegraphics[height=3 cm, width=4.2 cm, clip=true, angle=0]
{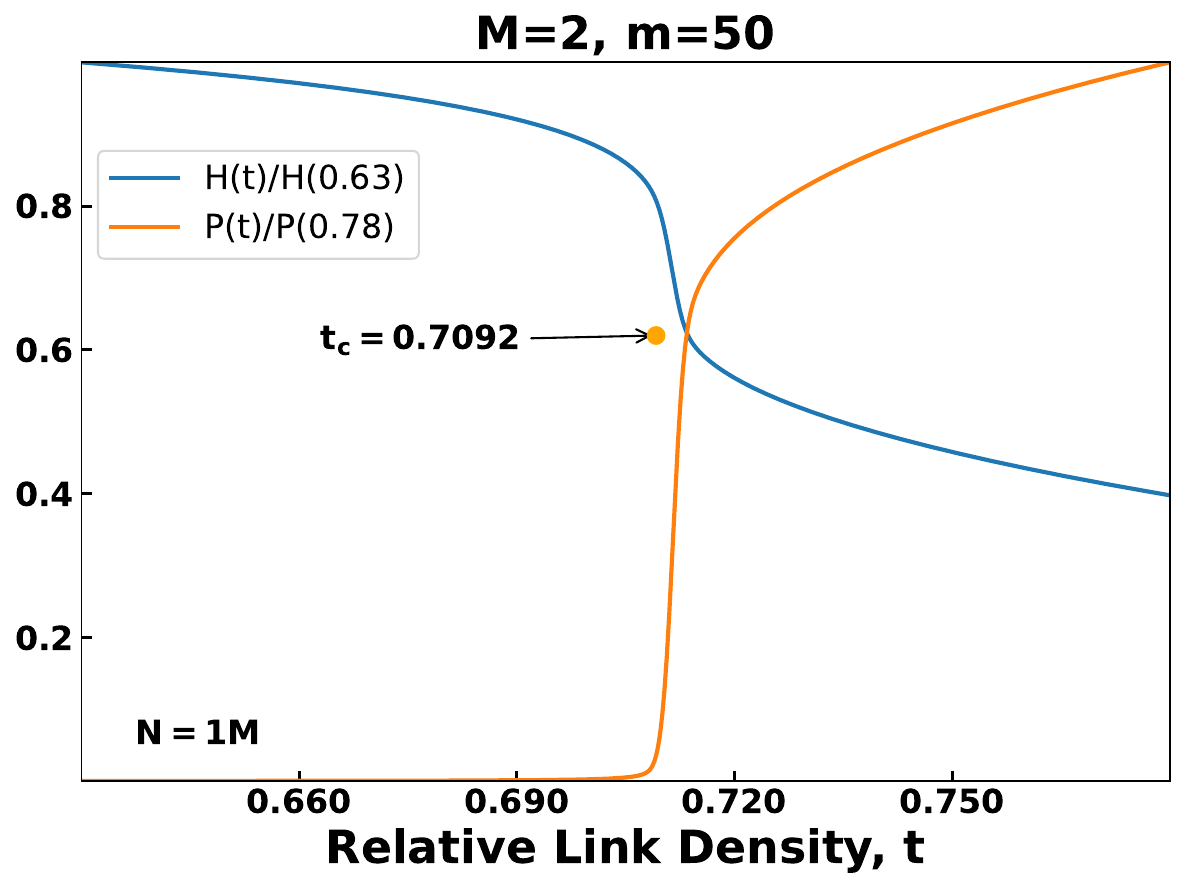}
\label{fig:3a}
}
\subfloat[]
{
\includegraphics[height=3 cm, width=4.2 cm, clip=true, angle=0]
{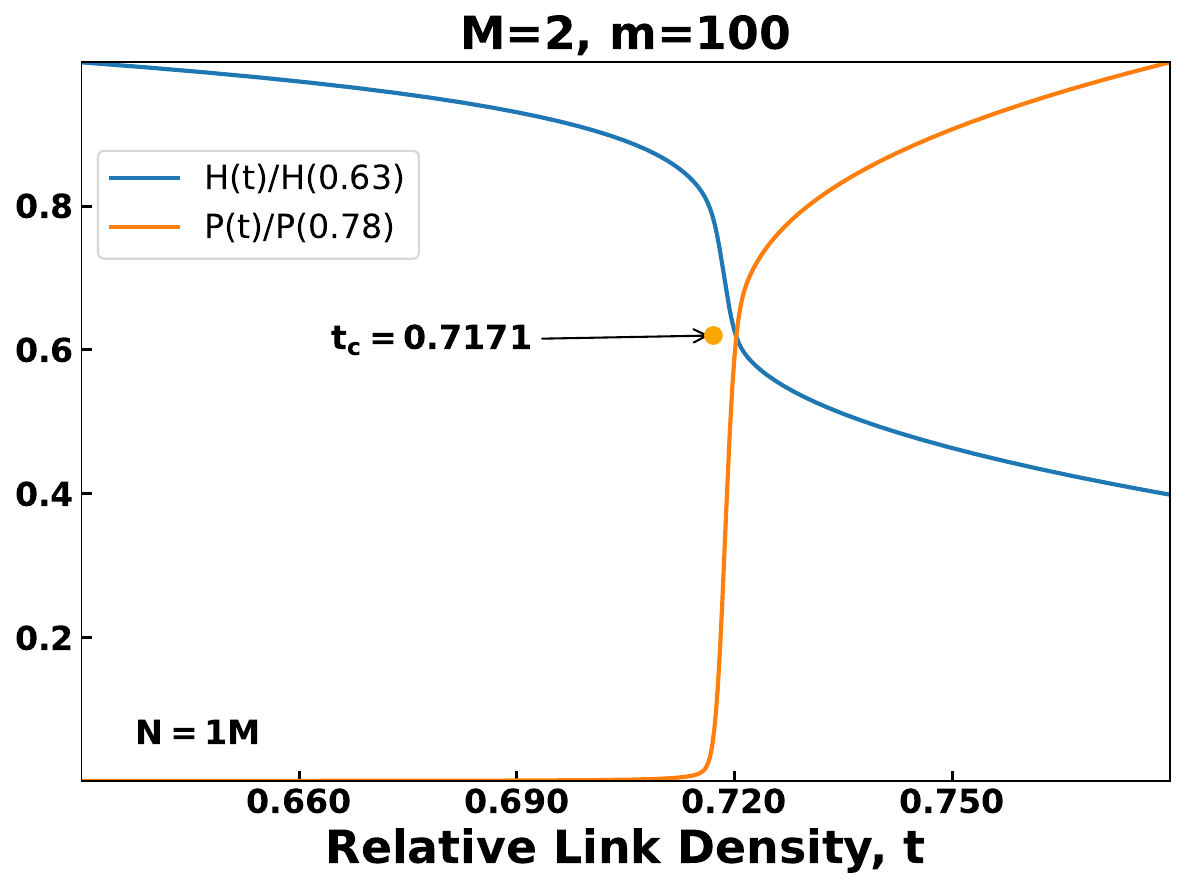}
\label{fig:3b}
}

\subfloat[]
{
\includegraphics[height=3 cm, width=4.2 cm, clip=true, angle=0]
{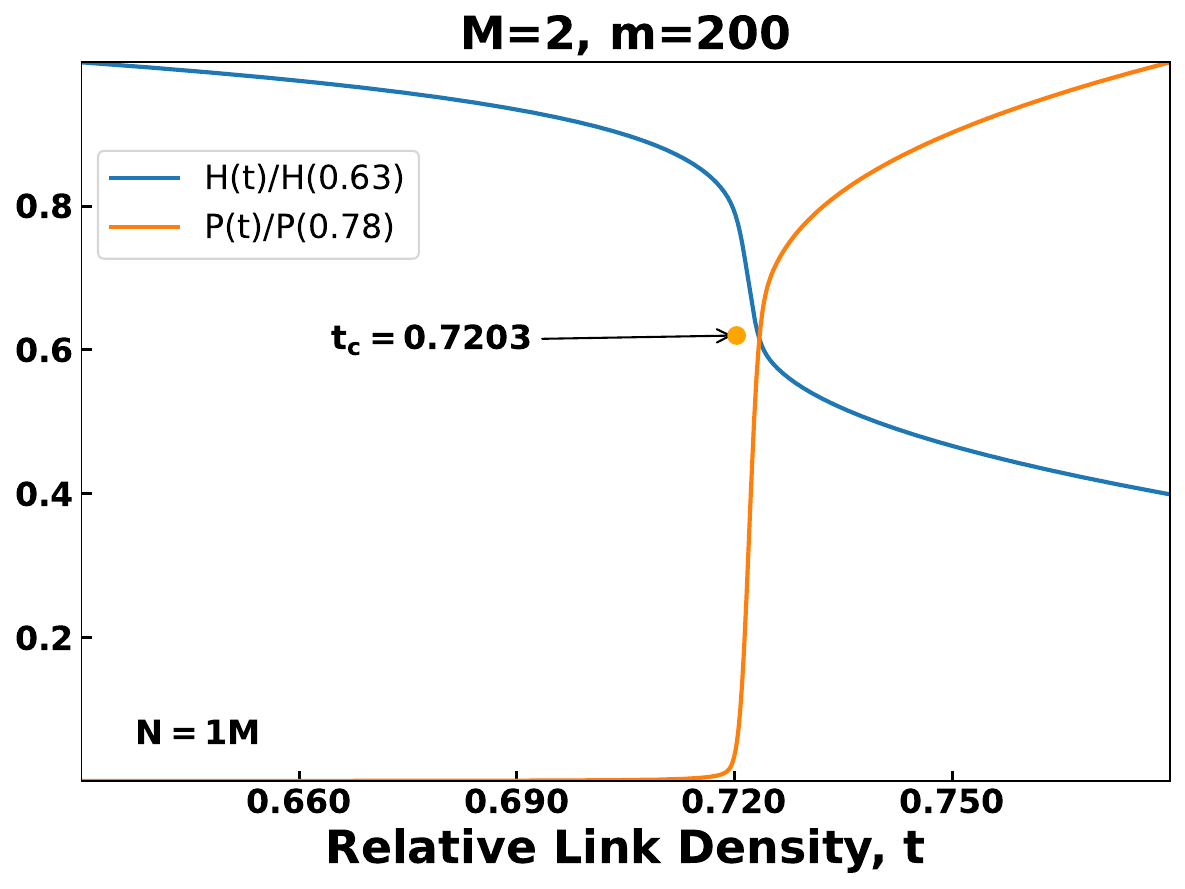}
\label{fig:3c}
}
\subfloat[]
{
\includegraphics[height=3 cm, width=4.2 cm, clip=true, angle=0]
{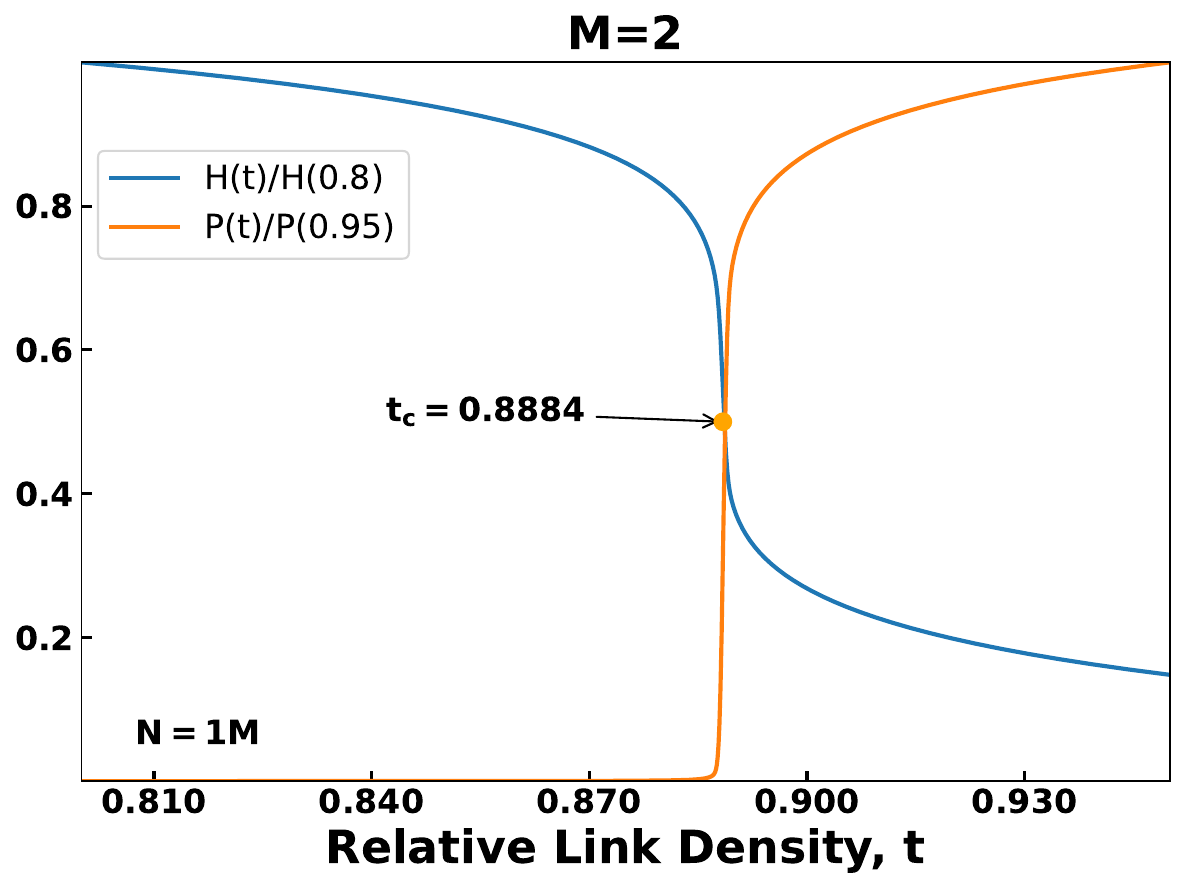}
\label{fig:3d}
}
\caption{Plots of relative entropy and relative order parameter for 
(a) $m=50,M=2$ (b) $m=100,M=2$ (c) $m=200,M=2$ and (d) presents the corresponding result for ER network with $M = 2$, shown for comparison.} 

\label{fig:3abcd}
\end{figure}
Both studies reported that, at the early stage of the process where the order parameter is zero, entropy also reaches its minimum value, seemingly contradicting the fundamental notion that a system cannot simultaneously be in the ordered and disordered phases. 

In 2017, Hassan et al. showed that this apparent contradiction arises from an inappropriate choice of probability in the definition of Shannon entropy.
Specifically, the summation in Shannon entropy was not taken over  
cluster sizes but over the number of clusters, and the probabilities used did not properly reflect the fraction of nodes contained in each cluster. They defined $\mu_i=s_i/N$ as the cluster picking probability that a site being picked at random will belong to the $i$th cluster \cite{ref.hassan_didar}.
We then used it in the definition of Shannon entropy 
\begin{equation}
\label{eq:shannon_entropy}
H(t)=-\sum_i^m \mu_i\log \mu_i,
\end{equation} 
where the summation is over every distinct labelled cluster $i=1,2,..., m$ \cite{ref.shannon}. 

It is worthwhile to explore the connection between Shannon entropy $H(t)$ and
Boltzmann entropy $S(t)= k \log \Omega(t)$. Suppose that at time $t$ 
the system consists of $m$ clusters with sizes $s_1, s_2, \dots, s_m$, such that $\sum_{i=1}^m s_i = N$.
The $N$ initially isolated nodes can be distinctly arranged into $m$ clusters in the
following number of ways:
\begin{equation}
\Omega = \frac{N!}{s_1! s_2! \cdots s_m!}.
\end{equation}
Hence, the Boltzmann entropy is
\begin{equation}
S = \log \frac{N!}{s_1! s_2! \cdots s_m!},
\end{equation}
where we set $k=1$ since it simply corresponds to a choice of entropy units.
Substituting $s_i = N\mu_i$ and applying Stirling’s approximation to the above expression, we obtain
\begin{equation}
S = HN.
\end{equation}
This leads to the relation $H = \frac{S}{N}$, which establishes a philosophical connection between the system and the observer.
The Boltzmann entropy $S$ is an \textit{extensive} quantity that quantifies the
physical disorder of the system, whereas the Shannon entropy $H = S/N$ reflects the
\textit{observer’s uncertainty per observation}.

We plot both the entropy and the order parameter on the same graph to examine how sharply they vary near the critical point. When these changes become sufficiently abrupt to produce a clear demarcation, the transition can be explicitly identified as an order–disorder transition: the system resides in a high-entropy, low-$P$ state for $t<t_c$ and in a low-entropy, high-$P$ state for $t>t_c$. To this end, we focus on a suitable interval $[t_1,t_2]$ around the critical point and normalize the entropy $H$ by its value at $t_1$ and the order parameter $P$ by its value at $t_2$.  Figs.~\ref{fig:2abcdef} and \ref{fig:3abcd} display the normalized entropy $H(t)/H(t_1)$ and the normalized order parameter $P(t)/P(t_2)$ as functions of $t$. From these figures, it is evident that the normalized order parameter remains close to zero while the normalized entropy stays near its maximum in one regime, with the opposite occurring in the other. This behavior clearly signals the presence of two distinct phases—an ordered phase and a disordered phase—separated by a critical point $t_c$. Such a transition is the hallmark of order–disorder phenomena and is characterized by symmetry breaking across the critical point. Moreover, $t_c$ increases as $M$ increases, indicating a delay in the onset of percolation, as can be clearly seen in Figs.~\ref{fig:2abcdef} and \ref{fig:3abcd}. For a fixed value of $M$, the critical point $t_c$ also increases slightly with increasing $m$. For example, with $M=3$, the critical thresholds for $m=50$, $m=100$, and $m=200$ were found to be approximately $0.8751$, $0.8792$, and $0.8807$, respectively, as shown in Fig.~\ref{fig:2abcdef}.

Interestingly, we observe that the curves of entropy and the order parameter intersect at a single point, denoted by $t_{\mathrm{HP}}$. The difference, defined as $\delta = t_{\mathrm{HP}} - t_c$, decreases with increasing $M$ for fixed $m$. For $M = 2$, this difference $\delta$ is significantly larger than for higher values of $M$. For instance, with $M = 2$ and $m = 100$, we find $\delta = 0.00321$ (see Fig.~\ref{fig:3abcd}), and entropy remains considerably higher in the supercritical region. In contrast, for $M = 3$, $\delta$ is already very close to zero, and for $M = 4$ it becomes almost negligible (see Fig.~\ref{fig:2abcdef}). This behavior stands in sharp contrast to bond percolation on Erd\H{o}s--R'enyi (ER) networks, where $\delta$ is nearly zero even for $M=2$ (see Fig.~\ref{fig:3d}). Moreover, in ER networks entropy decreases sharply across $t_c$, to such an extent that when compared with the corresponding order parameter it can effectively be regarded as the ordered phase—a behavior closely resembling the $M=4$ case in the MDA network. Taken together, these findings suggest that the Achlioptas rule with $M=2$ does not, by itself, guarantee explosive percolation. Rather, the nature of the percolation transition depends crucially on the topology of the underlying skeleton, {\it vis-à-vis}.


Our results imply that the percolation for $M = 2$ is definitely not explosive in nature. The nature of order parameter across  clearly suggest that it is the second order phase transition but it is accompanied by weakly order-disorder  transition. However, the case for $M = 3$ is definitely accompanied by order-disorder transition but whether it is explosive in nature is remain to investigate. Nonetheless, the case for $M=4$ is definitely explosive in character. Consequently, the parameter $\delta$ may serve as a useful quantitative indicator for characterizing the nature and sharpness of percolation transitions. Nonetheless, further investigation is required to draw definitive conclusions regarding the fundamental nature of these transitions. We know that the disordered phase corresponds to the high temperature phase and ordered phase corresponds to the low temperature phase. Figs.~\ref{fig:2abcdef} and \ref{fig:3abcd} clearly suggest that $T = 1 - t$ effectively plays the role of a temperature.


\subsection{Entropic origins of powder keg effect}

Why does selecting one link among $M$ randomly chosen candidate links---based on the criterion of forming the smallest resulting cluster---delay the percolation transition yet make it more abrupt? Empirically, we observe that increasing $M$ leads to progressively sharper transitions. To understand the underlying physics, we consider the Helmholtz free energy per site:\begin{equation}
F = E - T H,
\end{equation}
where $E$ denotes the internal energy and $H$ the entropy per site. Thermodynamics dictates that at any given stage, the equilibrium configuration corresponds to the minimum of $F$. At high temperatures (or, equivalently, at small $t$), the entropic contribution dominates $F$, and its minimum occurs when $H$ reaches its maximum. In this regime, the system initially resides in a high-temperature disordered phase, characterized by maximal entropy and a minimal order parameter.
 In percolation, the essential mechanism is the formation of progressively larger clusters through the addition of inter-cluster links, regardless of the specific rule employed. However, every added link inevitably decreases entropy. When $M$ candidate links are available, the most favorable choice is the one that minimizes this entropy reduction, thereby allowing the system to retain the highest possible entropy consistent with link addition. This corresponds to selecting the link that results in the smallest cluster growth.
Consider the entropy change: a large drop, say $\Delta H = -10$, adds a large positive contribution $-T \Delta H = 10T$ to $F$, which then requires a substantial compensatory decrease in internal energy to remain favorable. In contrast, a smaller drop, such as $\Delta H = -1$, contributes only $T$ to $F$, making it energetically more favorable. Thus, selecting the link that minimally reduces entropy is thermodynamically favored. This is precisely the mechanism implemented in the {\it best-of-$M$} model, which delays the transition but sharpens it once it occurs.

A subsequent question arises: Why does the so-called \textit{powder keg} effect \cite{ref.Friedman} emerge only when smaller clusters are preferentially grown, and not when larger clusters are favored?
The answer lies in the entropy-dominated regime. In the disordered phase, minimizing free energy requires maximizing entropy. As $M$ increases, the system lingers longer in this high-entropy state, delaying the onset of percolation and pushing the critical threshold $t_c$ to larger values. During this delay, the network accumulates a large number of clusters of nearly equal size, confined within a narrow size window. The greater the number of clusters and the narrower the window, the higher the entropy; when all clusters are of equal size, entropy reaches its maximum. For sufficiently large $M$, this band of similarly sized clusters becomes so densely populated that, just below the critical point, the system is primed for sudden release. Consequently, the addition of only a few links can abruptly merge many clusters, triggering a cascade of coalescences and producing a sharp entropy drop. This highly synchronized merging—an avalanche-like collapse—constitutes the \textit{powder keg} effect and underlies the explosive nature of the percolation transition in the best-of-$M$ model. By contrast, if larger clusters were always favored, the system would concentrate mass into a few dominant clusters early on, smoothing the transition and preventing such a critical buildup.

\section{Finite-size scaling and Critical exponents}\label{sec:fss}


Finding the critical exponents of the order parameter, susceptibility, and specific heat—denoted by $\beta$, $\gamma$, and $\alpha$, respectively constitutes one of the most important objectives of this work. If these exponents could be determined analytically, they would by default correspond to the behavior of an infinite system. In practice, however, neither experiments nor numerical simulations can be performed on truly infinite systems.
This limitation is addressed through the \emph{finite-size scaling (FSS) hypothesis}, which enables the extrapolation of critical behavior in the thermodynamic limit from data obtained on finite-sized systems. An observable $F(t,N)$ is said to obey finite-size scaling if it satisfies
\begin{equation}
\label{eq:fss_c}
F(t,N)\sim N^{a/\nu}\phi_{F}\left((t-t_c)N^{1/\nu}\right),
\end{equation}
where $\phi_{F}(z)$ is the universal scaling function associated with the observable $F$
\cite{ref.fss_1, ref.fss_2, ref.fss_3, ref.barbar}. The theoretical basis of finite-size scaling (FSS) rests on Buckingham's $\Pi$-theorem of dimensional analysis \cite{ref.hassan_santo}. According to this theorem, both $F(t,N)N^{-a/\nu}$ and the scaling variable $(t-t_c)N^{1/\nu}$ are dimensionless, implying that their numerical values are independent of measurement units, including system size. A powerful way to verify Eq.~(\ref{eq:fss_c}) is via data collapse: when curves of $F(t,N)$ versus $t$ for different $N$ are expressed in these dimensionless forms, they collapse onto a single universal curve. This approach provides a reliable means to extract critical exponents for infinite systems from finite-size data, as discussed in the following subsections. Here, we show the representative plots for the cases $M=3$ with $m=50$ and $M=4$ with $m=200$ as they capture necessary features for our analysis.

\begin{figure}

\centering

\subfloat[]
{
\includegraphics[height=3 cm, width=4.2 cm, clip=true, angle=0]
{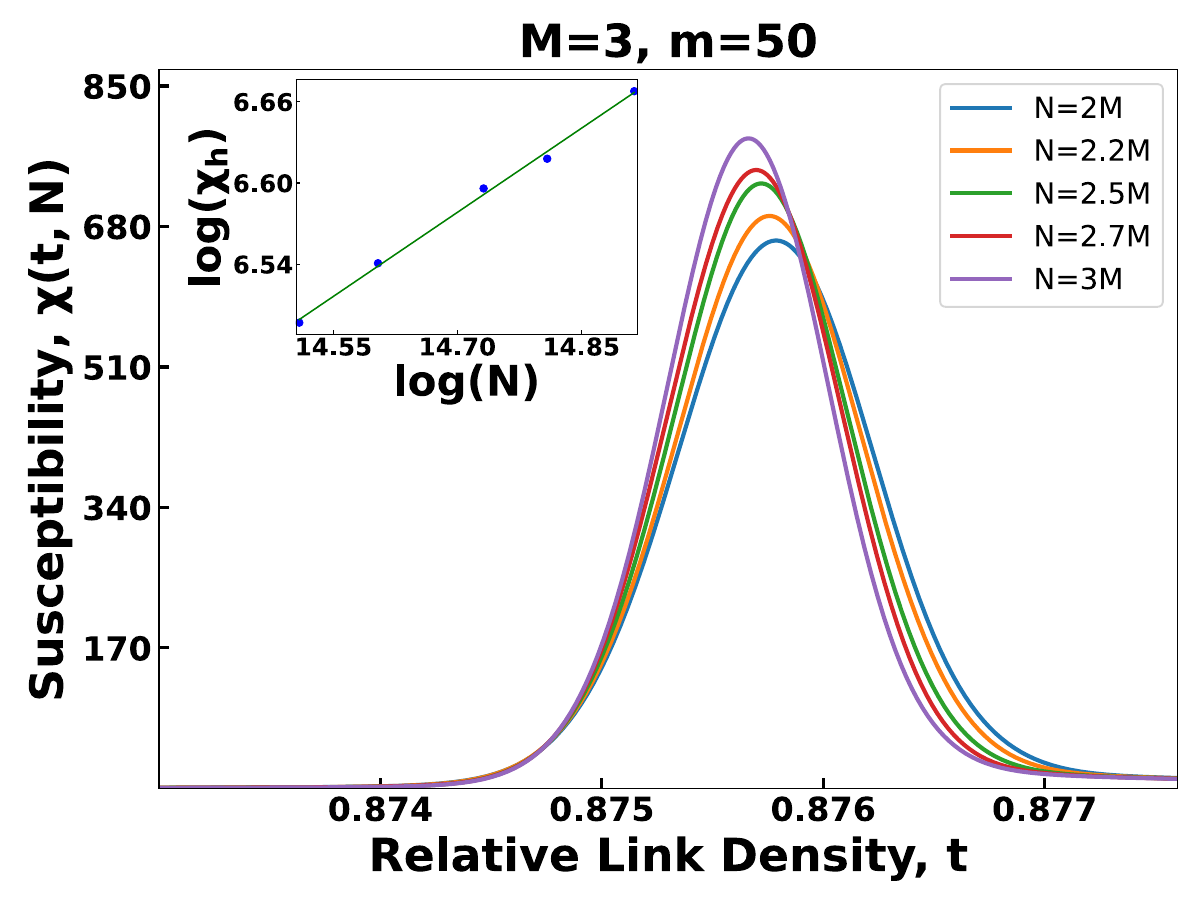}

\label{fig:4a}%
}
\subfloat[]
{
\includegraphics[height=3 cm, width=4.2 cm, clip=true, angle=0]
{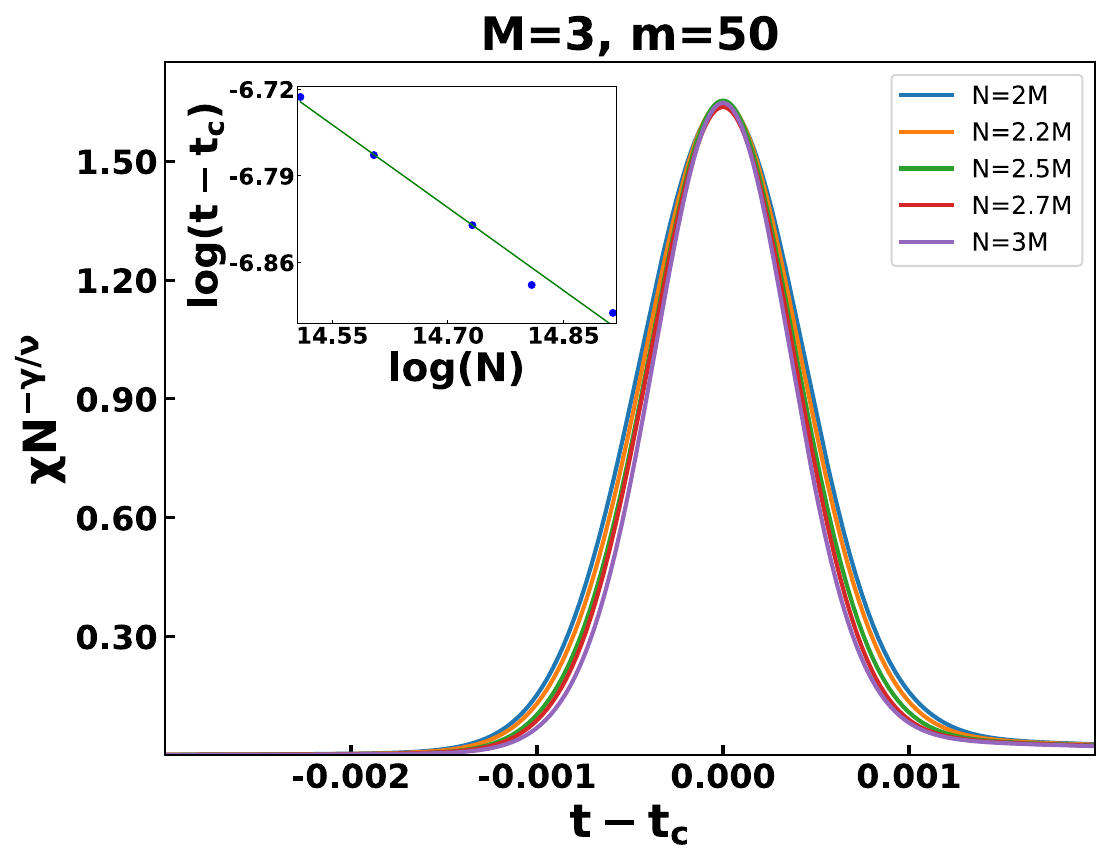}

\label{fig:4b}%
}

\subfloat[]
{
\includegraphics[height=3 cm, width=4.2 cm, clip=true, angle=0]
{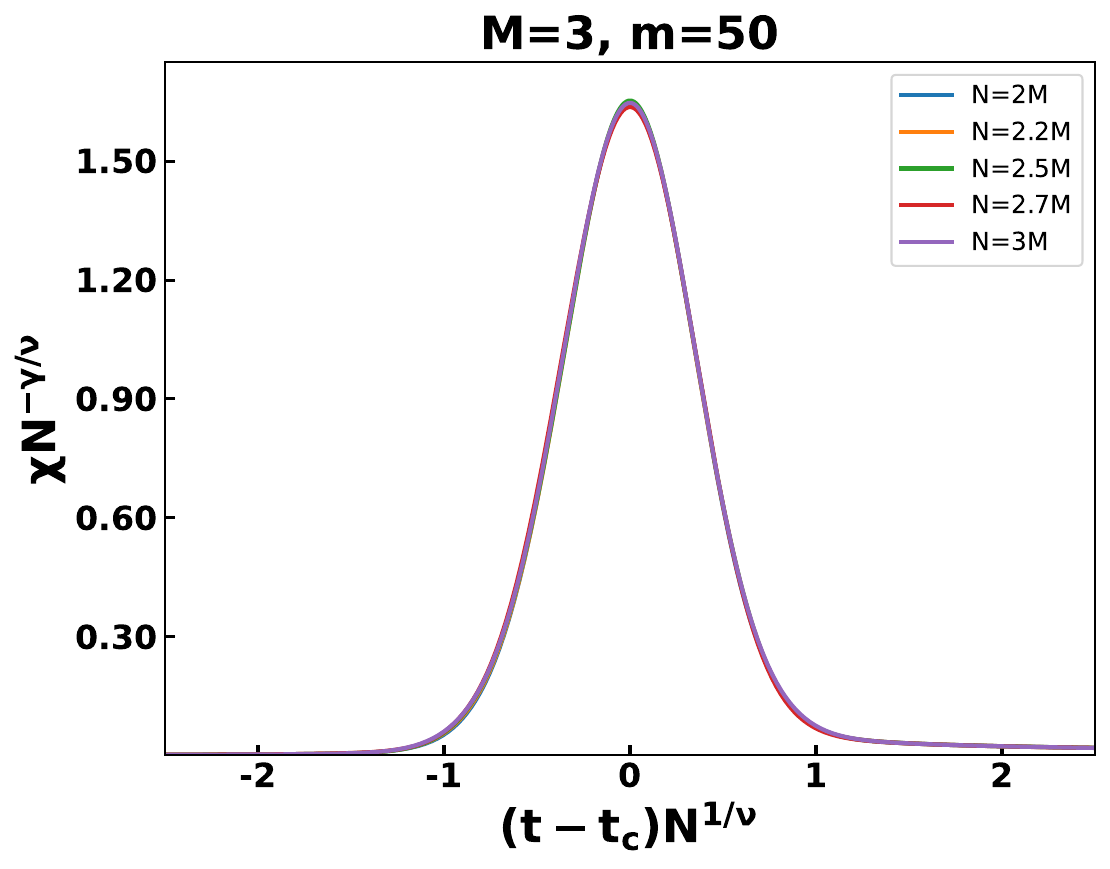}
\label{fig:4c}%
}
\subfloat[]
{
\includegraphics[height=3 cm, width=4.2 cm, clip=true, angle=0]
{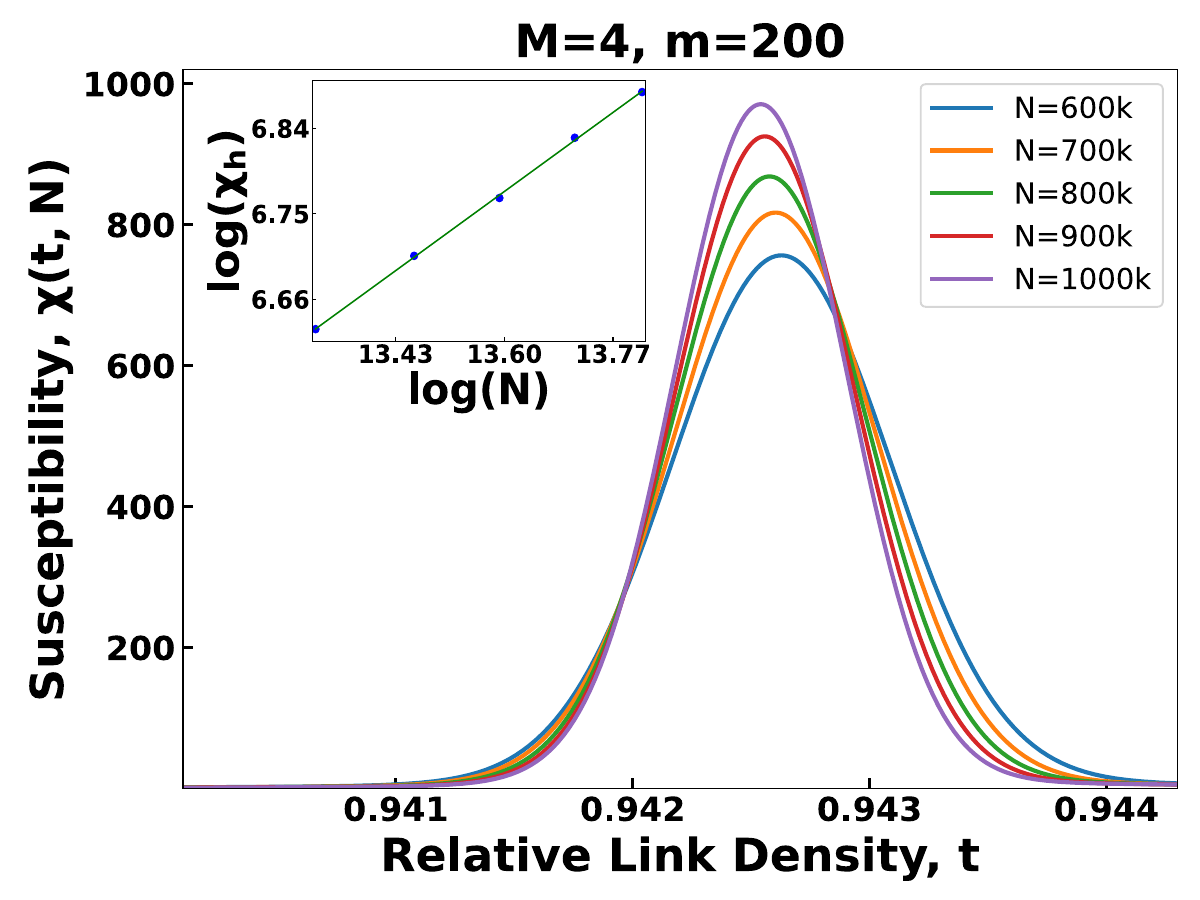}
\label{fig:4d}%
}

\subfloat[]
{
\includegraphics[height=3 cm, width=4.2 cm, clip=true, angle=0]
{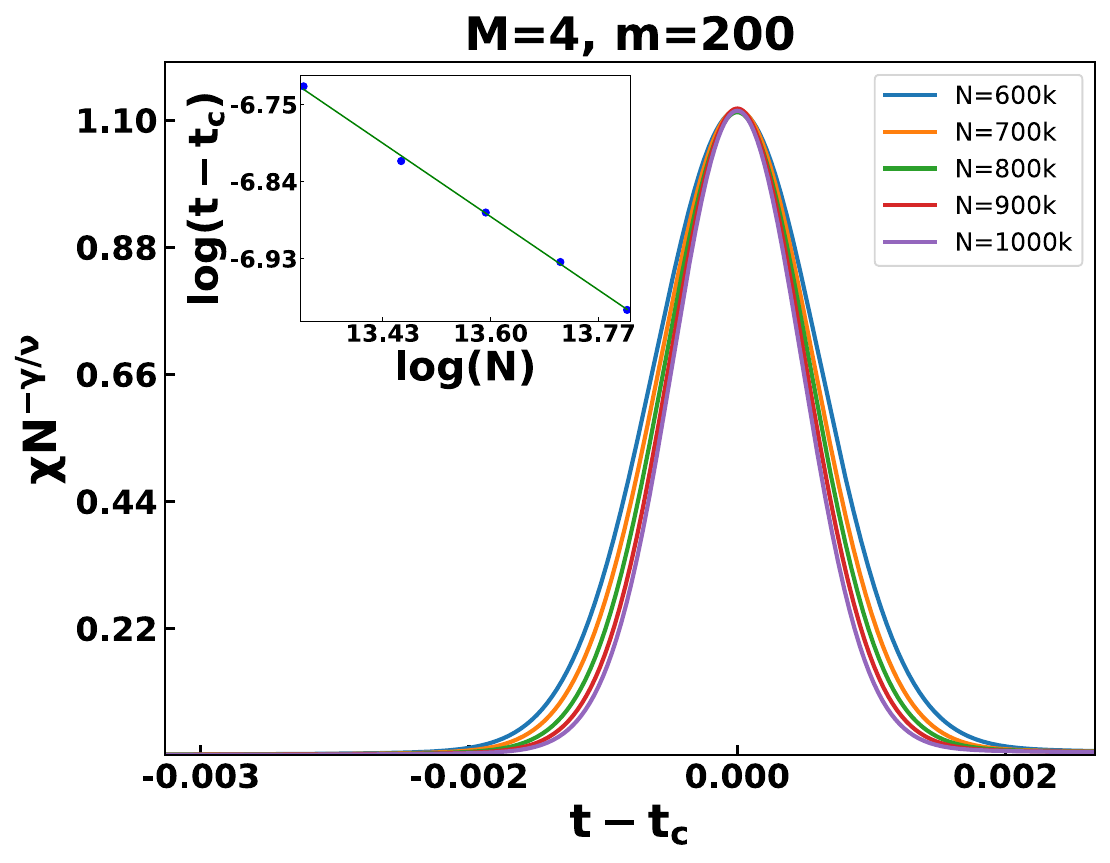}
\label{fig:4e}%
}
\subfloat[]
{
\includegraphics[height=3 cm, width=4.2 cm, clip=true, angle=0]
{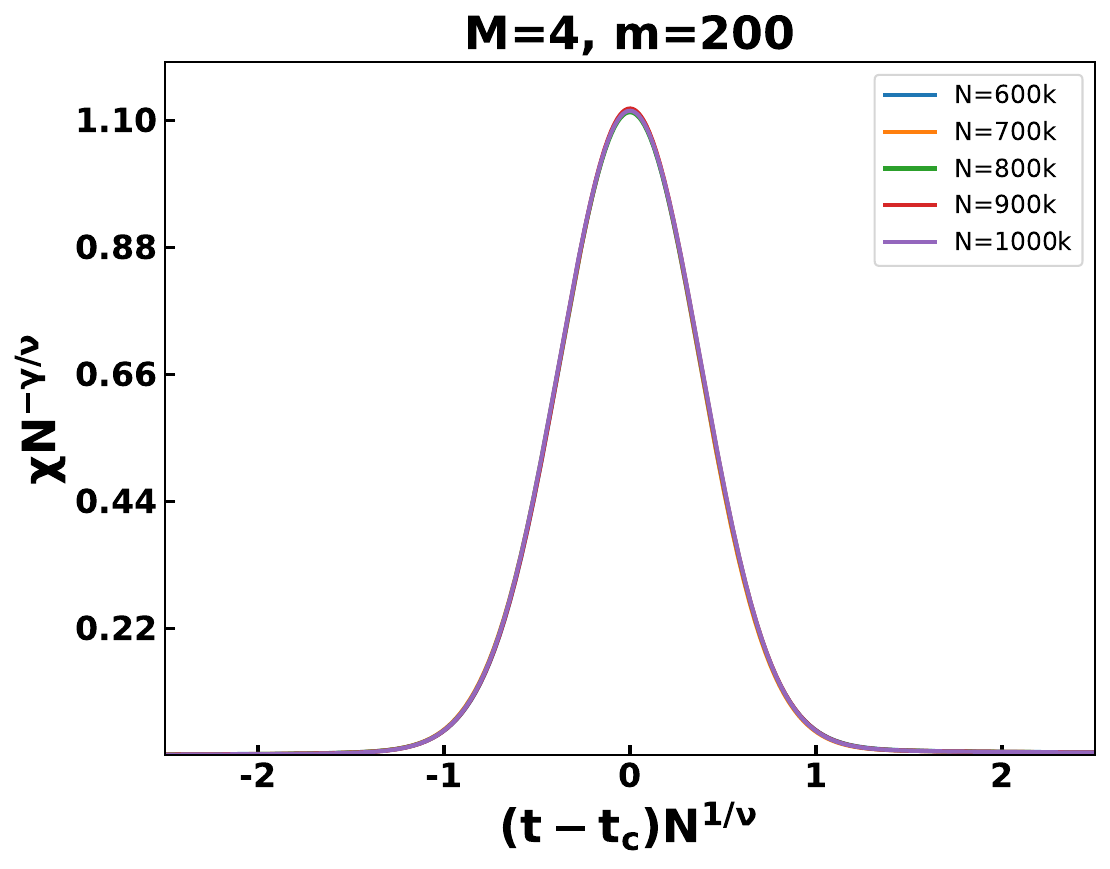}
\label{fig:4f}%
}

\caption{Plots of susceptibility $\chi$ versus relative link density $t$ for different
network sizes are shown in (a) and (d) for $M=3$ with $m=50$ and $M=4$ with $m=200$. In the inset we show
plots of $\log(\chi_h)$ versus $\log(N)$ and find straight lines whose slopes give an estimate of
$\gamma/\nu$. In (b) and (e) we plot $\chi(t,N)N^{-\gamma/\nu}$ versus $t-t_c(N)$ find that all the peaks
of the respective plot of (a) and (d) collapse at $t=t_c(N)$. The quality of peak collapse suggests
how good are the estimated values of $\gamma/\nu$. In the inset of (b) and (e) we show plots of 
$\log(t-t_c)$ versus $\log(N)$, the slopes of the resulting straight lines give an estimate
of $1/\nu$. When we plot $\chi(t,N)N^{-\gamma/\nu}$  versus $(t-t_c)N^{1/\nu}$ we find an 
excellent data collapse which proves that the values of $\gamma/\nu$ and $1/\nu$ are as good as the
theoretical values. 
} 

\label{fig:4abcdef}
\end{figure}

\subsection{Critical exponent of susceptibility}\label{subsec:fssA}

In the study of percolation, susceptibility is a central concept that quantifies the response of the system near the critical point. The notion that the second moment of the cluster size distribution $S(t,N)$ can serve as a measure of susceptibility was first introduced by Michael E. Fisher, who drew an analogy between percolation and thermodynamic critical phenomena \cite{ref.fisher}. Fisher observed that, much like the magnetic susceptibility in spin systems, the second moment of cluster sizes diverges near the percolation threshold, reflecting the emergence of large-scale connectivity. This insight has since become a cornerstone in understanding the critical behavior and scaling properties of percolation systems.
Note that this definition of susceptibility exhibits its expected divergence at the critical point only if the largest cluster is excluded from  the definition else it will continue to diverge. Furthermore $\gamma$ obtained by this method 
using 
\begin{equation}\label{eq:sec_moment}
    M_2(t,N) = \sum_{s} s^2 n_{s}(t) /N
\end{equation}
 is too large to obey the Rushbrooke inequality with positive critical exponent of the specific heat, where $n_s$ is the number of
clusters of size $s$ per site. These issues prompted Hassan {\it et al.} (2017) to redefine susceptibility more directly as the derivative of the order parameter. This redefinition resolves the inconsistencies by yielding a much smaller value of $\gamma$, sufficient to satisfy the Rushbrooke inequality~\cite{ref.hassan_didar}.

Specifically, Hassan {\it et al.} redefined the susceptibility as 
\begin{equation}\label{eq:susceptibility}
    \chi(t,N) = \frac{dP(t,N)}{dt}.
\end{equation}
In the discrete approximation, Eq. (\ref{eq:susceptibility}) becomes $\chi(t,N) \approx \Delta P(t,N)/\Delta t$ with $\Delta t = 1/N$ and hence in the thermodynamic limit it takes form as given in Eq. (\ref{eq:susceptibility}) \cite{ref.hassan_didar}.
Firstly, applying the FSS hypothesis to the susceptibility 
\begin{equation} 
\label{eq:fss_chi}
\chi(t,N)=N^{\gamma/\nu}\phi_\chi((t-t_c)N^{1/\nu}),
\end{equation}
where $\phi_\chi(z)$ is the scaling function for susceptibility. 
It clearly suggests that the peak height $\chi(t_c,N)=\chi_h$ at $t=t_c$ increases following a power-law
\begin{equation}
\label{eq:chi_h}
\chi_h\sim N^{\gamma/\nu}.
\end{equation}
 To verify this scaling, we plot $\log(\chi_h)$ versus $\log(N)$, as shown in the insets of Figs.~\ref{fig:4a} and \ref{fig:4d}, 
and observe a straight line whose slope yields an estimate of $\gamma/\nu$.  
We further normalize the susceptibility by plotting $\chi(t,N)/\chi_h$ versus $t-t_c(N)$, 
where all curves collapse at a single point, consistent with the fact that $\chi(t,N)/\chi_h$ is dimensionless.  
Through fine-tuning, we find that the best collapse occurs for $\gamma/\nu = 0.4136$ ($m=50, M=3$) 
and $\gamma/\nu = 0.4899$ ($m=200, M=4$).  

To estimate the exponent $\nu$, we locate the spread among the distinct curves 
and drop a perpendicular line from the common collapse point of the peaks.  
Measuring the distance from this intercept to each curve provides $(t-t_c)$ as a function of $N$.  
Plotting $\log(t-t_c)$ versus $\log(N)$ (see the insets of Figs.~\ref{fig:4b} and \ref{fig:4e}) yields a straight line, 
whose slope gives an estimate of $1/\nu$.  
We obtain $1/\nu = 0.4464$ for $m=50, M=3$ and $1/\nu = 0.5037$ for $m=200, M=4$.  
Finally, plotting $\chi(t,N)N^{-\gamma/\nu}$ as a function of $(t-t_c)N^{1/\nu}$ 
demonstrates that all distinct $\chi(t,N)$ versus $t$ curves collapse onto a universal scaling function, 
as shown in Figs.~\ref{fig:4c} and \ref{fig:4f}.  
The collapse occurs because both $\chi(t,N)/\chi_h$ and $(t-t_c)N^{1/\nu}$ are dimensionless quantities.  
Here, $1/\nu$ is the only tunable parameter, since $\gamma/\nu$ is already fixed.  
By combining $(t-t_c)\sim N^{-1/\nu}$ with the scaling relation $\chi_h\sim N^{\gamma/\nu}$, 
we arrive at
\begin{equation}
\chi(t,N)\sim (t-t_c)^{-\gamma},    
\end{equation}
with $\gamma = 0.9265$ for $m=50, M=3$ and $\gamma = 0.9726$ for $m=200, M=4$. 

\subsection{Critical exponent of specific heat}\label{subsec:fssB}
In 1961, Fisher proposed treating the number of clusters per site as the analog of the free energy $F$ in percolation \cite{ref.fisher}. By analogy with thermodynamics, he defined the specific heat as the second derivative of this quantity with respect to the link density $t$. Applying this definition to percolation on the square lattice yields the critical exponent $\alpha = -2/3$ \cite{ref.mertens_ziff}. According to thermodynamics, however, the first derivative of $F$ should correspond to entropy. 
In 2017, our attempt to define entropy opened the way to a more consistent formulation of specific heat, through direct mapping onto its thermodynamic definition:
\begin{equation}
\label{eq:sp_heat}
C(t,N) = (1-t)\frac{dH(t,N)}{d(1-t)},
\end{equation}
where $1-t$ serves as the effective analog of temperature, since the low-temperature phase must correspond to a disordered phase \cite{ref.hassan_didar}.
\begin{figure}[h]

\centering

\subfloat[]
{
\includegraphics[height=3 cm, width=4.2 cm, clip=true, angle=0]
{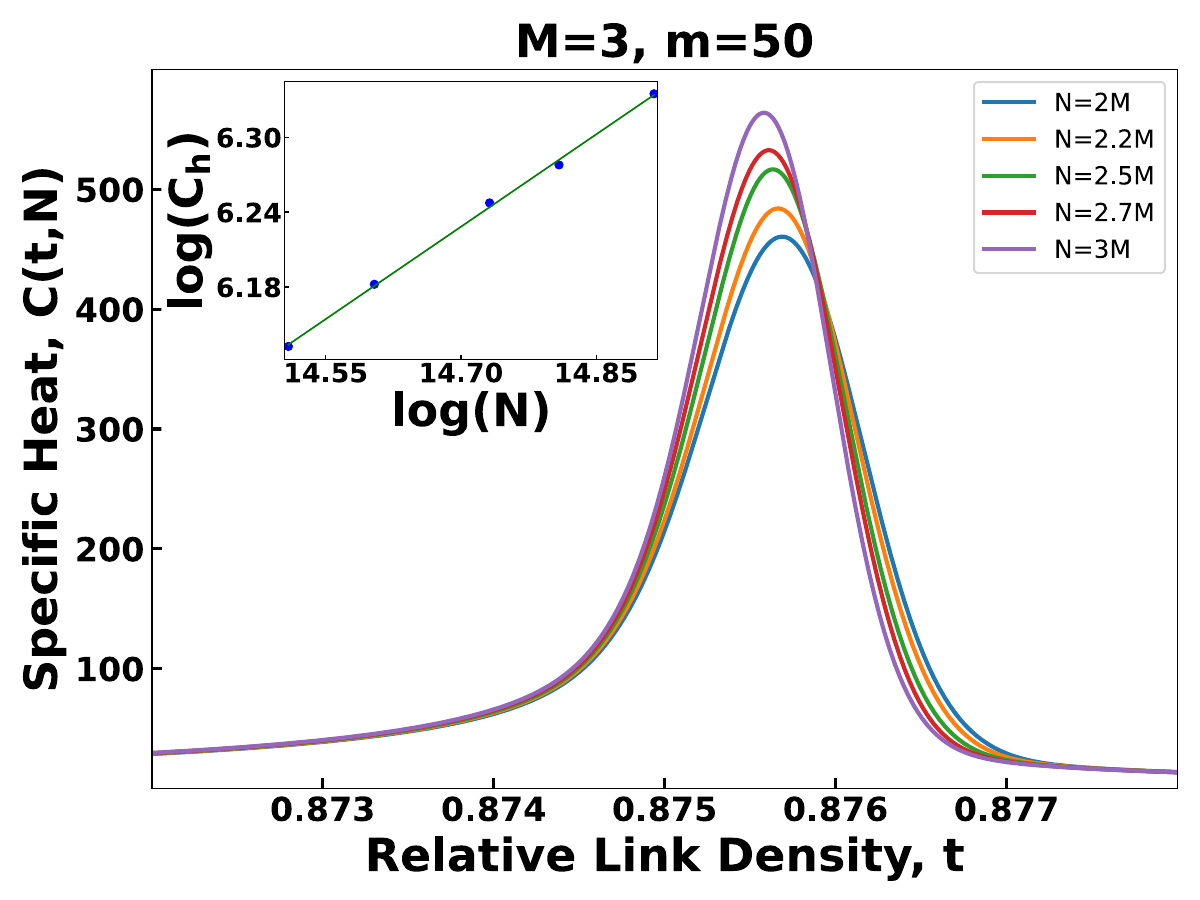}
\label{fig:5a}%
}
\subfloat[]
{
\includegraphics[height=3 cm, width=4.2 cm, clip=true, angle=0]
{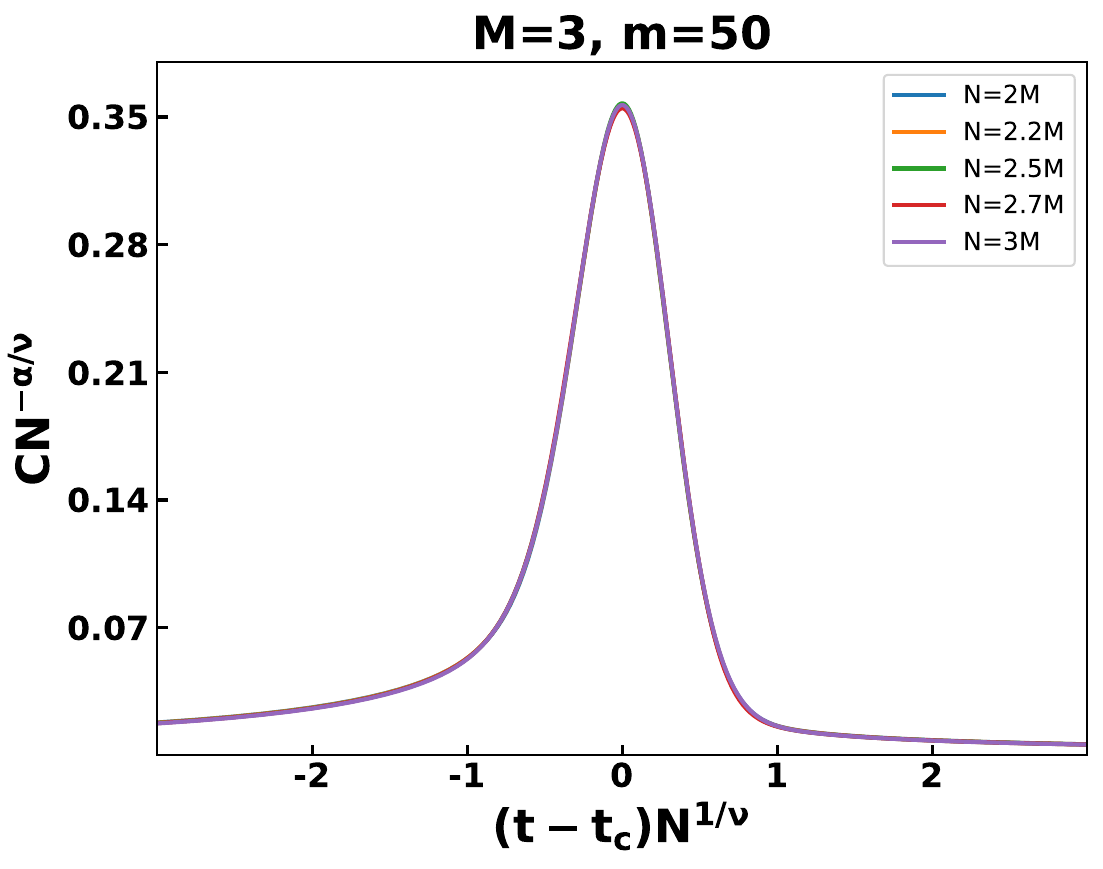}
\label{fig:5b}%
}

\subfloat[]
{
\includegraphics[height=3 cm, width=4.2 cm, clip=true, angle=0]
{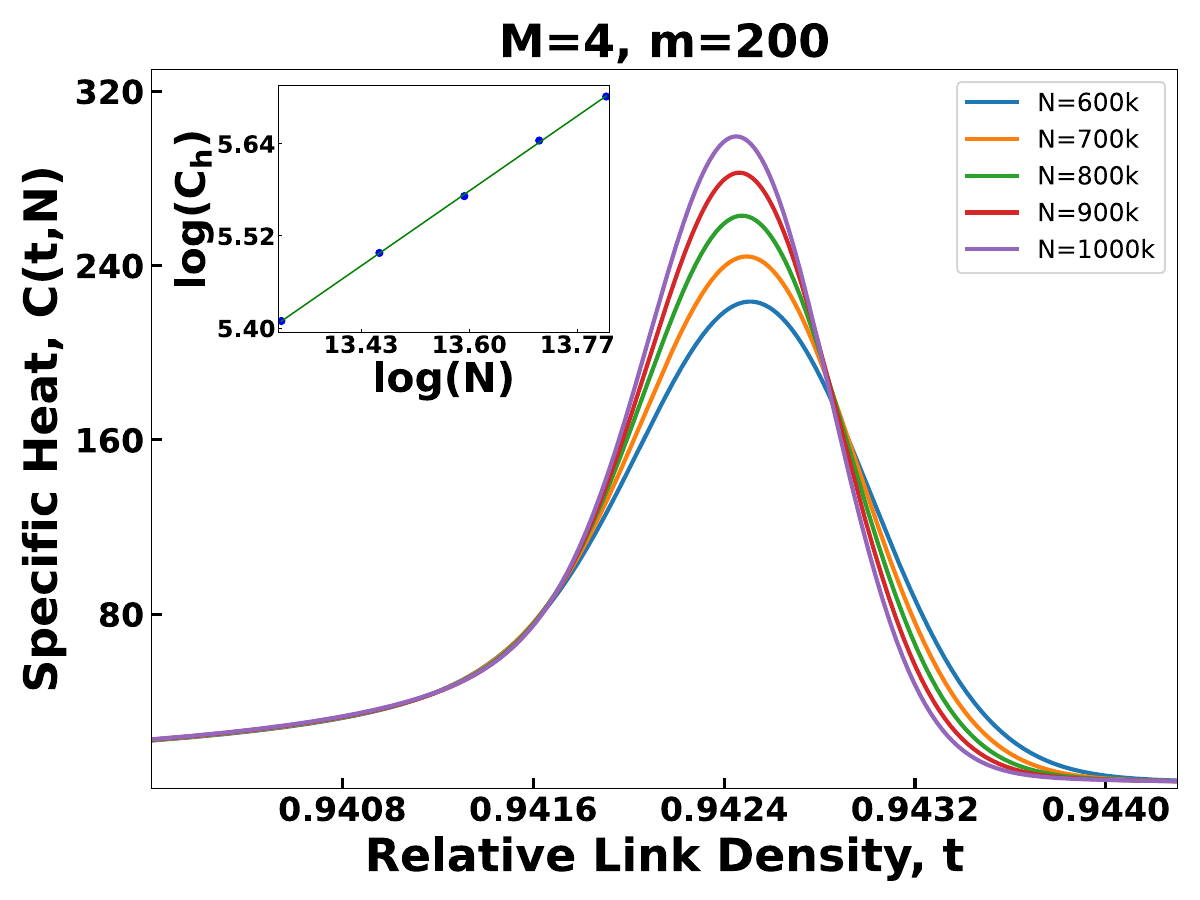}
\label{fig:5c}%
}
\subfloat[]
{
\includegraphics[height=3 cm, width=4.2 cm, clip=true, angle=0]
{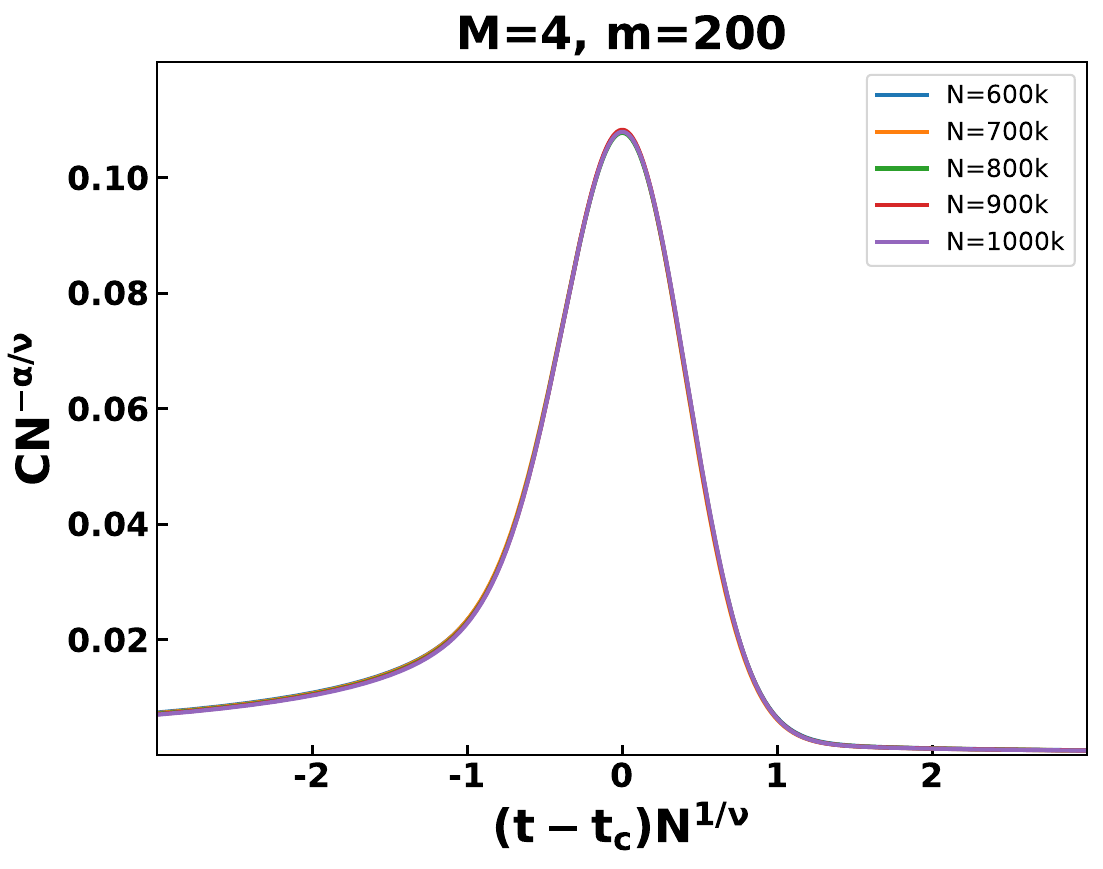}
\label{fig:5d}%
}
\caption{Plots of specific heat $C(t,N)$ versus $t$ for different
network sizes are shown in (a) and (c) for $M=3$ with $m=50$ and $M=4$ with $m=200$. In the inset we show
plots of $\log(C_h)$ versus $\log(N)$ and find straight lines whose slopes give an estimate of 
$\alpha/\nu$. We used the $1/\nu$ which was previously found from susceptibility for corresponding $M$ and $m$. Then in (b) and (d) we plot 
$C(t,N)N^{-\alpha/\nu}$  versus $(t-t_c)N^{1/\nu}$ and obtain an excellent data collapse revealing
that the values of $\alpha/\nu$ and $1/\nu$ are the best we can get numerically.
} 

\label{fig:5abcd}
\end{figure}
 Applying this definition to the square lattice, we obtained a positive value of $\alpha$, in contrast to Fisher’s $\alpha = -2/3$. A positive $\alpha$ carries significant implications: it indicates that the specific heat diverges at the critical point in a manner similar to susceptibility, rather than exhibiting a cusp or a gentle kink when $\alpha=-2/3$. This not only alters the classification of the critical singularity but also provides a deeper thermodynamic consistency to the percolation transition.

Using the definition given in Eq. (\ref{eq:sp_heat}), we measured the specific heat, which is shown in Figs.~\ref{fig:5a} and \ref{fig:5c} for $m=50, M=3$ and for $m=200, M=4$, respectively. Clearly, its behavior is almost identical to that of the susceptibility. To determine the corresponding $\alpha$ value, we employ the following finite-size scaling (FSS) hypothesis:
\begin{equation}
\label{eq:fss_specific}
C(t,N)\sim N^{\alpha/\nu}\phi_C\left((t-t_c)N^{1/\nu}\right).
\end{equation}
Following the same procedure as for the susceptibility, we calculate the exponent $\alpha/\nu$ for different values of $M$ and $m$. For instance, we obtain $\alpha/\nu=0.4939$ for $m=50, M=3$ and $\alpha/\nu=0.5738$ for $m=200, M=4$. Since the value of $\nu$ is already known, we now plot $C(t,N)N^{-\alpha/\nu}$ as a function of $(t-t_c)N^{1/\nu}$ and observe that all the distinct plots of Figs.~\ref{fig:5a} and \ref{fig:5c} collapse onto a single universal curve, as shown in Figs.~\ref{fig:5b} and \ref{fig:5d}, for $m=50, M=3$ and $m=200, M=4$, respectively.
As before, by using $(t-t_c)\sim N^{-1/\nu}$ in the relation $C_h\sim N^{\alpha/\nu}$, we immediately obtain
\begin{equation}
C(t,N)\sim (t-t_c)^{-\alpha},
\end{equation}
where $\alpha=1.1064$ for $m=50, M=3$ and $\alpha=1.1392$ for $m=200, M=4$. The quality of the data collapse once again serves as a litmus test of the accuracy of the estimated critical exponent $\alpha$. It is noteworthy that, similar to the square lattice case, we find $\alpha>0$ in the MDA network as well. It suggests that like in the continuous thermal phase transition, specific heat in percolation too diverges near the critical point following a power-law.%

\subsection{Critical exponent of order parameter}

\begin{figure}

\centering

\subfloat[]
{
\includegraphics[height=3 cm, width=4.2cm, clip=true, angle=0]
{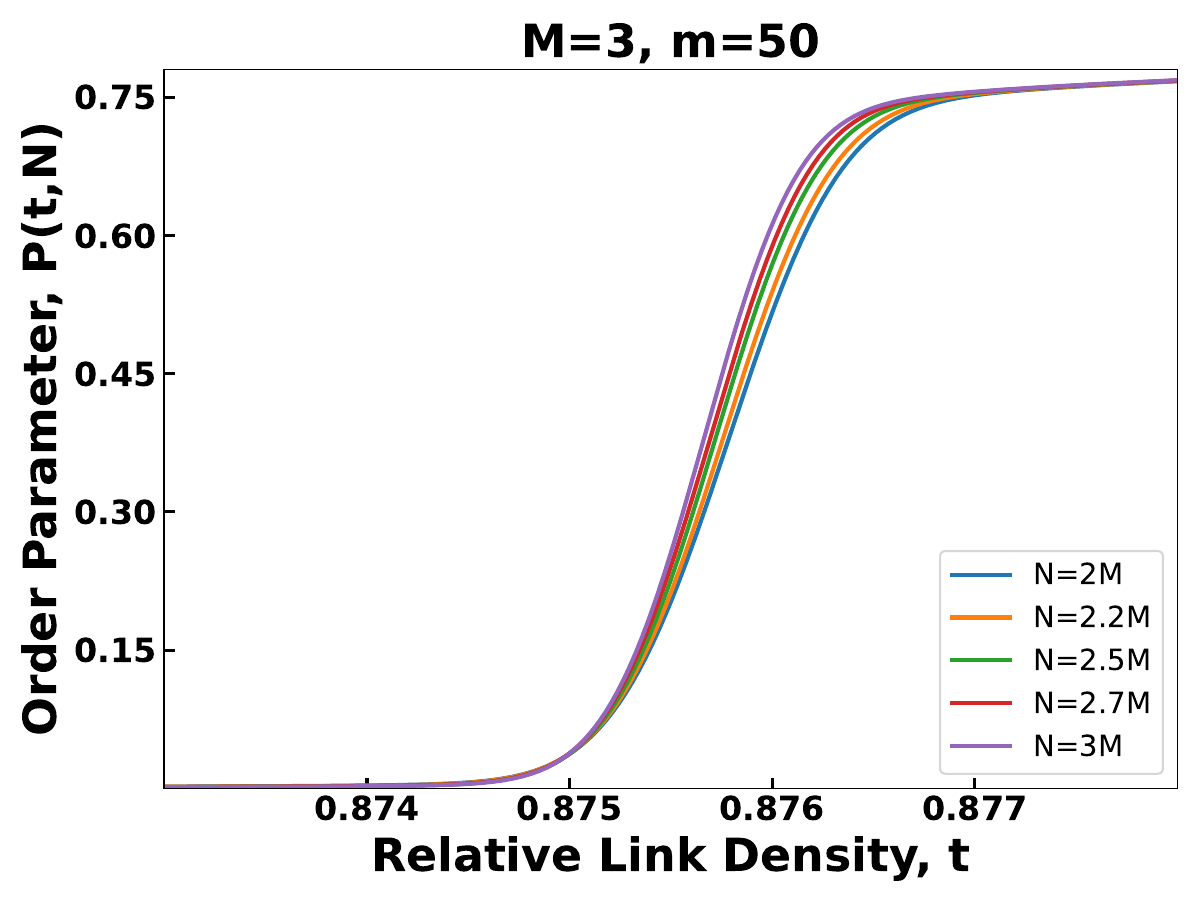}
\label{fig:6a}%
}
\subfloat[]
{
\includegraphics[height=3 cm, width=4.2 cm, clip=true, angle=0]
{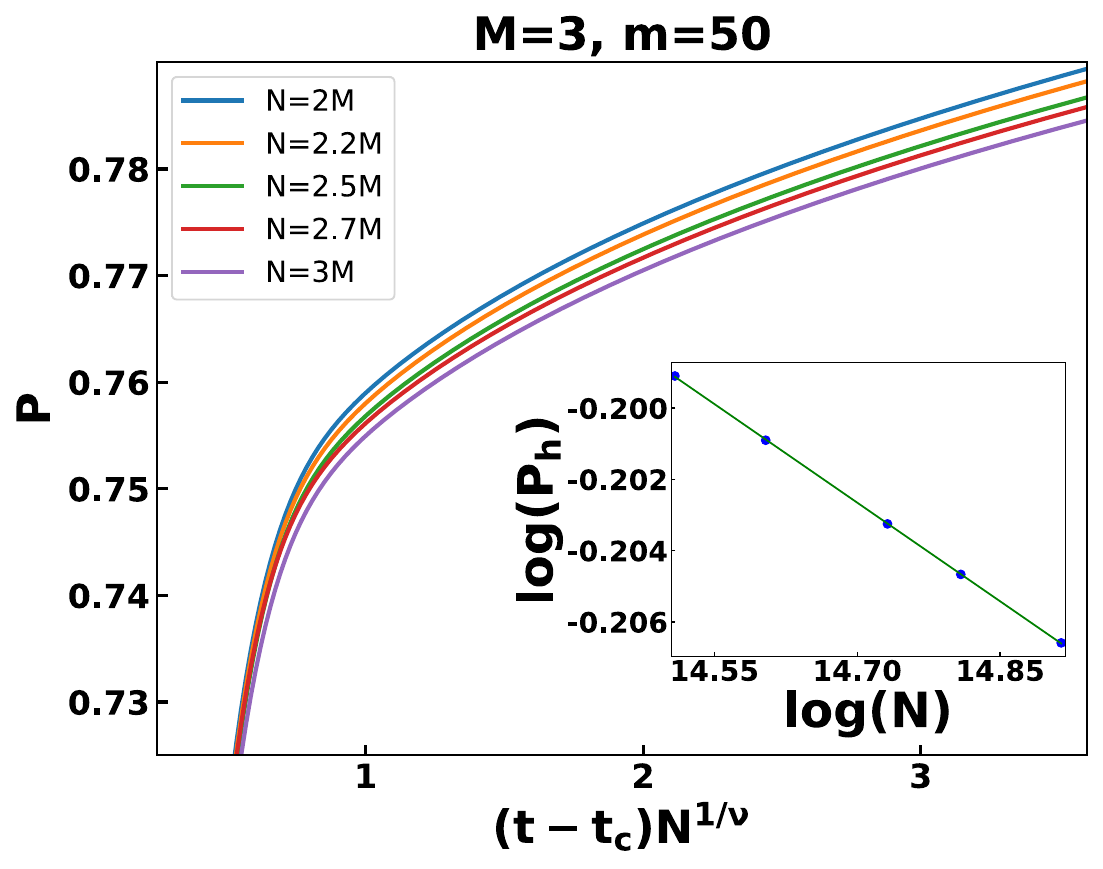}
\label{fig:6b}%
}

\subfloat[]
{
\includegraphics[height=3 cm, width=4.2 cm, clip=true, angle=0]
{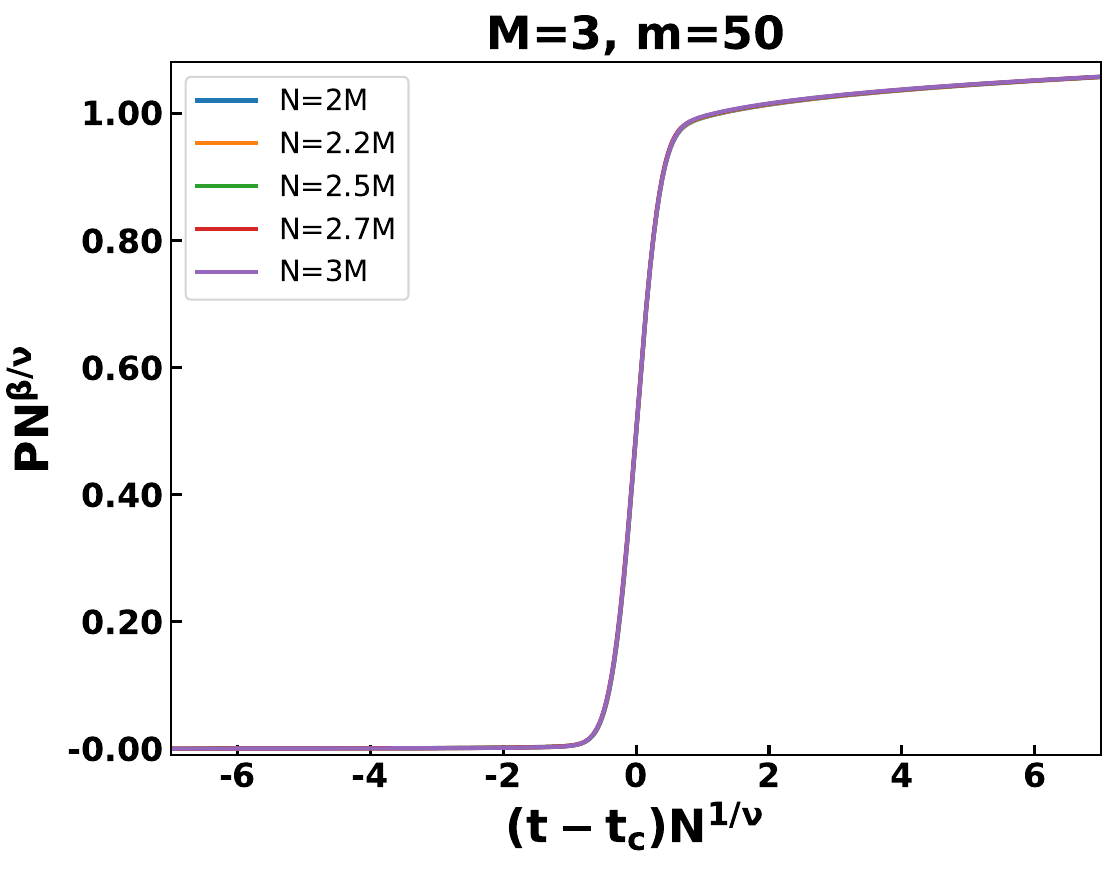}
\label{fig:6c}%
}
\subfloat[]
{
\includegraphics[height=3 cm, width=4.2 cm, clip=true, angle=0]
{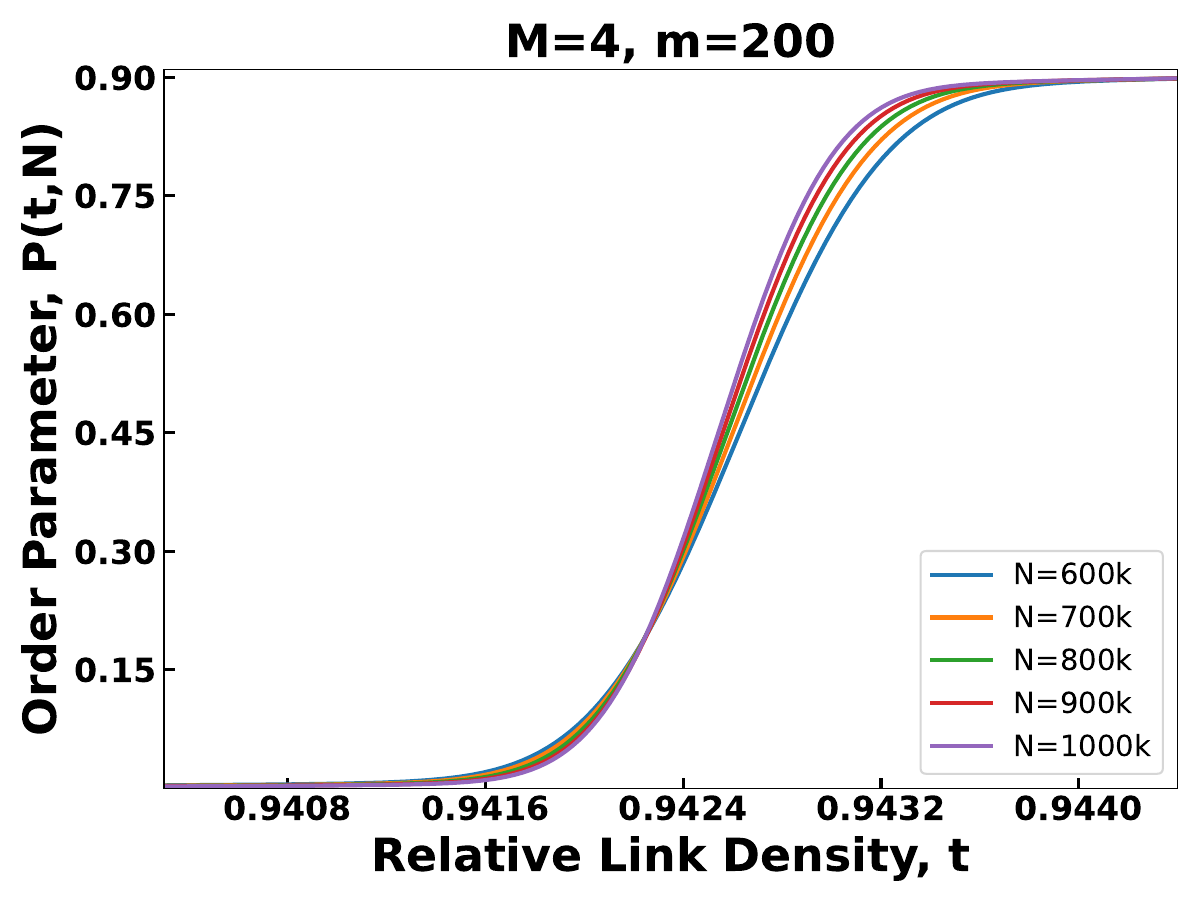}
\label{fig:6d}%
}

\subfloat[]
{
\includegraphics[height=3 cm, width=4.2 cm, clip=true, angle=0]
{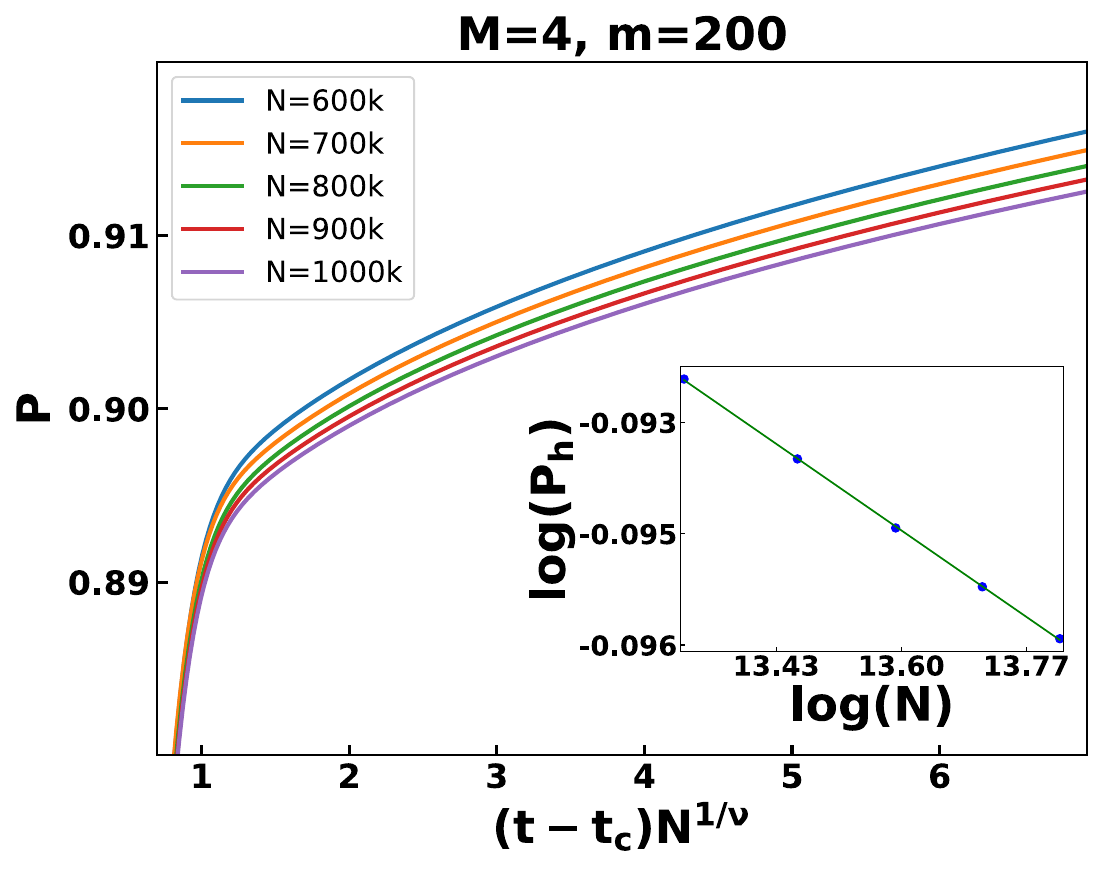}
\label{fig:6e}%
}
\subfloat[]
{
\includegraphics[height=3 cm, width=4.2 cm, clip=true, angle=0]
{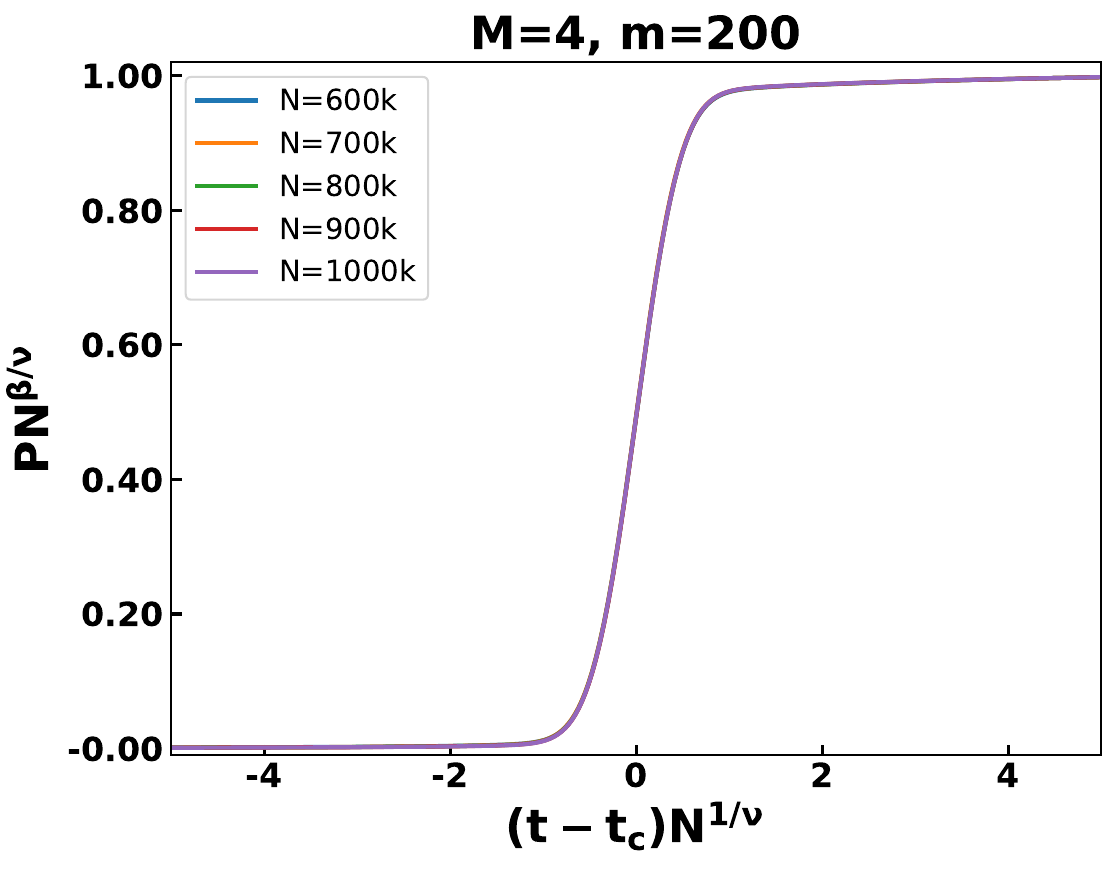}
\label{fig:6f}%
}

\caption{Plots of order parameter $P$ versus relative link density $t$ drawn in (a) and (d) for $m=50$, $M=3$ and $m=200$, $M=4$ for different network sizes $N$. $P$ versus $(t-t_c)N^{1/\nu}$ is plotted in (b) and (e) for for $m=50$, $M=3$ and $m=200$, $M=4$ respectively. In the insets of (b) and (e) $\log(P_h)$ versus
$\log(N)$ are shown which are straight lines with slopes $\beta/\nu$. Finally, in (c) and (f)
we plot $P(t,N)N^{\beta/\nu}$ versus $(t-t_c)N^{1/\nu}$ and finding excellent data collapse revealing
that the values of $\beta/\nu$ and $1/\nu$ are as close as the theoretical ones. 
} 

\label{fig:6abcdef}
\end{figure}

The order parameter in percolation was also first introduced by Fisher and Essam in 1961 and they defined it as the relative size of largest cluster $P(t,N)=s_{{\rm max}}/N$ which is found to behave like magnetization in magnetic systems. Later, Kasteleyn and Fortuin (1969–1972), through the random cluster model, showed that percolation corresponds to the $q \to 1$ limit of the Potts model. These contributions firmly established $P(t,N)$ as the standard order parameter in percolation theory. Order parameter $P$ of percolation takes the typical sigmoidal shape if we plot it
as a function of $1-t$ instead $t$ since $1-t$ is the equivalent counterpart of temperature, not the link density $t$ (see for details in \cite{ref.digonto}). 

The order parameter $P(t,N)$ is known
to obey the finite-size scaling 
\begin{equation}
P(t,N)\sim N^{-\beta/\nu}\phi_P((t-t_c)N^{1/\nu}),
\end{equation}
where $\phi_P(x)$ is the universal scaling function of $P$. Since we already know that $\nu$
value, we can immediately plot $P(t,N)$ versus $(t-t_c)N^{1/\nu}$. We see from the resulting plot 
that the height $P_h$ of the peak decreases with 
$N$. Measuring $P_h$ as a function of $N$ and plotting $\log(P_h)$ versus $\log(N)$ once again gives a straight line, see the insets of Figs.~\ref{fig:6b} and \ref{fig:6e}. First observation is that the slope is negative and the slope gives the value of $\beta/\nu$. Now, plotting $P(t,N)N^{\beta/\nu}$ versus $(t_c-t)N^{1/\nu}$ and 
tuning the value of $\beta/\nu$ we find excellent data collapse, see Figs.~\ref{fig:6c} and 
\ref{fig:6f} for $m=50,M=3$ and $m=200,M=4$ respectively. Using $P_h\sim N^{-\beta/\nu}$ in the relation $(t-t_c)\sim N^{-1/\nu}$ yields 
\begin{equation}
    P(t,N)\sim (t-t_c)^\beta, 
\end{equation}
where $\beta=0.0414$  for $m=50, M=3$, $\beta=0.0135$ for $m=200, M=4$. This relation is indeed reminiscent of the magnetization in the paramagnetic to ferromagnetic transition. 

\section{Universality}

In the study of second-order phase transitions, the concept of \textit{universality} captures the remarkable fact that systems with vastly different microscopic details can exhibit identical macroscopic behavior near criticality. 
First introduced by Kadanoff in 1970, universality asserts that systems sharing the same critical exponents and scaling functions fall into the same universality class, irrespective of their underlying microscopic structures. Not only this, the critical exponents must also obey some scaling relation like Rushbrooke inequality which reduces to equality under the static scaling hypothesis~\cite{ref.Stanley}. We have already obtained critical exponents for different $M$ values while 
keeping $m$ fixed and vice versa. We observed that for a fixed network topology characterized by fixed $m$, increasing $M$ delays the onset of percolation; however, once the transition begins, it becomes markedly more abrupt. This trend is captured in the decreasing values of the order-parameter exponent $\beta$ with increasing $M$, as shown in Table~\ref{tab:critical_expo}, indicating a steeper rise of the order parameter near the percolation threshold. In contrast, the critical exponents $\gamma$ and $\alpha$, associated with susceptibility and specific heat respectively, increase with $M$. This reinforces the increasingly explosive character of the transition as $M$ increases.

Further insight emerges when the network topology is varied while keeping the number of choices $M$ fixed. For example, with $M=3$ and $M=4$, the percolation transition remains explosive across all examined topologies. In these cases, as $m$ increases from $50$ to $200$, the exponent $\beta$ decreases, while $\gamma$ and $\alpha$ increase, consistent with the trends observed earlier for fixed topology and varying $M$. However, for $M=2$ on MDA networks, the transition in the supercritical regime is not perfectly ordered, as the entropy remains relatively high even above $t_c$. A similar behavior is observed for $M=2$ in the case of percolation on the Barabási–Albert (BA) network. 
It is important to note that the critical exponents obey the Rushbrooke inequality, $\alpha + 2\beta + \gamma \geq 2$, for $M=2,3$ and $M=4$ and for varying $m$ such as $m=50, 100$ and $m=200$. 
\begin{table}[h]
    \centering
   
    \begin{tabular}{|c|c|c|c|c|c|c|c|}
        
        \hline
        $M$ & $m$ & $t_c$ & $\nu$ & $\alpha$ & $\beta$ & $\gamma$ & $\alpha+2 \beta+\gamma$ \\
        \hline
        \hline
        \multirow{3}*{2}     & 50 & 0.7092 & 2.1505 & 1.0297 & 0.1166 & 0.9101 & 2.173  \\ \cline{2-8}
                            & 100 & 0.7171 & 2.0534 & 1.0179 & 0.1162 & 0.8961 & 2.1464 \\ \cline{2-8}
                            & 200 & 0.7203 & 1.8619 & 0.9538 & 0.1156 & 0.8695 & 2.0545 \\ 
        \hline
        \hline
        \multirow{3}*{3}     & 50 & 0.8751 & 2.2401 & 1.1064 & 0.0414 & 0.9265 & 2.1157 \\ \cline{2-8}
                            & 100 & 0.8792 & 1.9877 & 1.1127 & 0.0358 & 0.9686 & 2.1529 \\ \cline{2-8}
                            & 200 & 0.8807 & 1.9861 & 1.1156 & 0.0338 & 0.969  & 2.1522 \\ 
        \hline
        \hline
        \multirow{3}*{4}     & 50 & 0.9391 & 2.0534 & 1.1066 & 0.0154 & 0.9417 & 2.0791 \\ \cline{2-8}
                            & 100 & 0.9414 & 1.9562 & 1.124  & 0.0145 & 0.9724 & 2.1254 \\ \cline{2-8}
                            & 200 & 0.9422 & 1.9853 & 1.1392 & 0.0135 & 0.9726 & 2.1388 \\ 
        \hline
        
    \end{tabular}
     \caption{Values of critical thresholds, exponents, and Rushbrooke inequalities for explosive percolation on MDA networks for different $m$ and $M$ combinations.
}
    \label{tab:critical_expo}
\end{table}
In all cases, the Rushbrooke inequality is satisfied, with near-equality observed, further validating the robustness of the scaling relations in these explosive percolation transitions. Also note that, the critical exponents depend only weakly on $m$ but exhibit a strong dependence on 
$M$, indicating that the universality classes differ not only across different $M$ but also for different $m$, unlike BA networks. When $m$ is varied from 50 to 200, the degree exponent $\omega$ exhibits a small shift, from 2.9358 to 2.9989, implying that the universality depends weakly on the degree exponent or network topology.

\section{Summary}

We have studied bond percolation on mediation-driven attachment (MDA) networks under the generalized Achlioptas process. In this framework, multiple candidate bonds are sampled, and the one minimizing the resulting cluster size is selected according to the best-of-$M$ rule. This competitive growth mechanism systematically suppresses the formation of large clusters. At the same time, it promotes the growth of smaller ones, leading to nontrivial percolation dynamics that depend on both the degree exponent $\omega$, governed by the parameter $m$, and the choice parameter $M$.

Our results show that the critical threshold $t_c$ and the associated critical exponents $(\beta, \alpha, \gamma)$ vary jointly with $\omega$ and $M$. While earlier studies have established that explosive percolation can occur on Erdős–Rényi (ER) for as few as two options ($M=2$), we find that in MDA networks the transition becomes explosively continuous only when $M \geq 3$. For $M=2$, the transition is second order in nature but lacks a well-defined order–disorder transition, which we can regard as a hybrid-type transition. This difference highlights the essential role of degree heterogeneity and mediation-driven growth in shaping the percolation transition.

An important finding is that the universality class of the transition depends not only on $M$ but also on the degree exponent $\omega$. This is in sharp contrast to percolation on BA networks, where universality classes remain 
invariant under changes in $m$. The weak dependence of the exponents on $m$ and the strong dependence on $M$ in the MDA networks indicate that the choice mechanism is the dominant driver of critical behavior in MDA networks, where the ``winner-takes-all'' effect is almost washed out. The percolation on MDA networks, where the 'winner-takes-all' effect is strong, requires special care, which we shall address in our future endeavors.


To better characterize the nature of the transition, we performed a joint analysis of the entropy $H$ and the order parameter $P$. Plotting these two quantities on the same scale reveals a striking diagnostic feature: the distance $\delta$ between the critical point $t_c$ and the point where the entropy and order parameter curves meet and remain closely aligned. When $\delta \approx 0$ and the two curves stay nearly coincident over a finite range, the transition is unequivocally explosive and proceeds via symmetry breaking, as observed for $M=3$ and $M=4$ across all values of $m$ we have studied. In contrast, for $M=2$, $\delta$ remains finite, and the two curves merely intersect rather than staying aligned, indicating that the transition is not even weakly explosive but rather a weak second-order transition and that too without symmetry breaking. This metric therefore serves as a simple yet powerful tool for distinguishing between conventional continuous and explosive second-order transitions. Furthermore, the Rushbrooke inequality, $\alpha + 2\beta + \gamma \geq 2$, was found to be consistently satisfied. However, the behavior of $\alpha$ and $\gamma$ for $M=2$ remains ambiguous and warrants further investigation.

For higher values of $M$, the transition becomes markedly sharper. In particular, when $M=4$, the order parameter exhibits a distinctly abrupt jump compared to the case of $M=3$. This pronounced sharpening can be attributed to an enhanced powder-keg effect: as macroscopic cluster formation is systematically suppressed, a vast number of mesoscopic clusters accumulate throughout the system. When the system approaches the critical threshold $t_c$, these intermediate clusters rapidly merge, triggering a sudden, large-scale coalescence resulting in the sharp drop in entropy.  This mechanism underscores the entropic origin of explosive percolation. The deliberate suppression of early large-cluster formation preserves a high-entropy configuration for a longer duration. However, once the system reaches its tipping point, entropy-driven coalescence initiates a dramatic, seemingly discontinuous drop in entropy—signaling the abrupt emergence of the giant component.

Given the ubiquity of scale-free topologies in real-world systems such as communication, transportation, and biological networks, understanding how percolation unfolds on MDA-type networks is of both theoretical and practical importance \cite{ref.clote,ref.masucci,ref.lou,ref.briesemeister,ref.cui,ref.kosmidis,ref.tian,ref.kitsak,ref.wu_gao}. Our findings reveal that the explosive character of percolation transitions is 
sensitive to both the underlying degree distribution and the competitive rules of bond selection. This sensitivity suggests that real-world networks may be tuned—either toward robustness or fragility—by structural interventions that mimic the role of $m$ and $M$.

\begin{acknowledgments}
NR, MMBS, and MKH thank the BdREN of the University Grants Commission (UGC) of Bangladesh for providing excellent computing facilities.
\end{acknowledgments}

\end{document}